\documentclass{aastex63}

\usepackage[caption=false]{subfig}

\newcommand{\source}{\object[HB93 1156+295 Ton599 4C~+29.45]{1156+295}}
\newcommand{\compstar}{\object{HD105199}}

\newcommand{\FeLower}{2250}
\newcommand{\FeUpper}{2750}

\newcommand{\NRmag}{599}
\newcommand{\MinR}{18.4}
\newcommand{\MaxR}{13.4}
\newcommand{\RObsStart}{2014 December 9}
\newcommand{\RObsEnd}{2019 June 3}
\newcommand{\ObsGammaEnd}{2019 September 6}
\newcommand{\NGamma}{584}
\newcommand{\NGammaUpper}{127}

\submitjournal{The Astrophysical Journal}
\accepted{25 December 2021}

\begin{document}

\title{Emission Line Variability during a Nonthermal Outburst in the Gamma-Ray Bright Quasar 1156+295}

\author[0000-0002-6634-9673]{Melissa K. Hallum}
\affiliation{Institute for Astrophysical Research, Boston University,
    725 Commonwealth Ave, Boston, MA 02215}
\email{mhallum@bu.edu}
\author[0000-0001-6158-1708]{Svetlana G. Jorstad}
\affiliation{Institute for Astrophysical Research, Boston University,
    725 Commonwealth Ave, Boston, MA 02215}
\affiliation{Sobolev Astronomical Institute, St. Petersburg State University, St. Petersburg, Russia}
\author[0000-0002-4640-4356]{Valeri M. Larionov}
\altaffiliation{Deceased}
\affiliation{Sobolev Astronomical Institute, St. Petersburg State University, St. Petersburg, Russia}
\affiliation{Pulkovo Observatory, St.-Petersburg, 196140, Russia}
\author[0000-0001-7396-3332]{Alan P. Marscher}
\affiliation{Institute for Astrophysical Research, Boston University,
    725 Commonwealth Ave, Boston, MA 02215}
\author[0000-0003-1134-7352]{Manasvita Joshi}
\affiliation{Institute for Astrophysical Research, Boston University,
    725 Commonwealth Ave, Boston, MA 02215}
\affiliation{Research Computing, Information Technology Services, Northeastern University }
\author[0000-0001-6314-0690]{Zachary R. Weaver}
\affiliation{Institute for Astrophysical Research, Boston University,
    725 Commonwealth Ave, Boston, MA 02215}
\author[0000-0003-1318-8535]{Karen E. Williamson}
\affiliation{Institute for Astrophysical Research, Boston University,
    725 Commonwealth Ave, Boston, MA 02215}
\author[0000-0002-3777-6182]{Iv\'an Agudo}
\affiliation{Instituto de Astrof\'{\i}sica de Andaluc\'{\i}a,  CSIC, Apartado 3004, 18080 Granada, Spain}
\author{George A. Borman}
\affiliation{Crimean Astrophysical Observatory RAS, P/O Nauchny, 298409, Russia}
\author[0000-0003-1117-2863]{Carolina Casadio}
\affiliation{Foundation for Research and Technology, Hellas, IESL and Institute of Astrophysics, Voutes, 7110 Heraklion, Greece}
\affiliation{Department of Physics, University of Crete, 71003, Heraklion, Greece}
\author[0000-0002-8773-4933]{Antonio Fuentes}
\affiliation{Instituto de Astrof\'{\i}sica de Andaluc\'{\i}a,  CSIC, Apartado 3004, 18080 Granada, Spain}
\author{Tatiana S. Grishina}
\affiliation{Sobolev Astronomical Institute, St. Petersburg State University, St. Petersburg, Russia}
\author[0000-0001-9518-337X]{Evgenia N. Kopatskaya}
\affiliation{Sobolev Astronomical Institute, St. Petersburg State University, St. Petersburg, Russia}
\author[0000-0002-2471-6500]{Elena G. Larionova}
\affiliation{Sobolev Astronomical Institute, St. Petersburg State University, St. Petersburg, Russia}
\author{Liyudmila V. Larionova}
\affiliation{Sobolev Astronomical Institute, St. Petersburg State University, St. Petersburg, Russia}
\author[0000-0002-9407-7804]{Daria A. Morozova}
\affiliation{Sobolev Astronomical Institute, St. Petersburg State University, St. Petersburg, Russia}
\author[0000-0001-9858-4355]{Anna A. Nikiforova}
\affiliation{Sobolev Astronomical Institute, St. Petersburg State University, St. Petersburg, Russia}
\author[0000-0003-4147-3851]{Sergey S. Savchenko}
\affiliation{Sobolev Astronomical Institute, St. Petersburg State University, St. Petersburg, Russia}
\affiliation{Special Astrophysical Observatory, Russian Academy of Sciences, 369167, Nizhnii Arkhyz, Russia}
\affiliation{Pulkovo Observatory, St.-Petersburg, 196140, Russia}
\author[0000-0002-4218-0148]{Ivan S. Troitsky}
\affiliation{Sobolev Astronomical Institute, St. Petersburg State University, St. Petersburg, Russia}
\author[0000-0002-9907-9876]{Yulia V. Troitskaya}
\affiliation{Sobolev Astronomical Institute, St. Petersburg State University, St. Petersburg, Russia}
\author[0000-0002-8293-0214]{Andrey A. Vasilyev}
\affiliation{Sobolev Astronomical Institute, St. Petersburg State University, St. Petersburg, Russia}

\begin{abstract}
    We present multi-epoch optical spectra of the $\gamma$-ray bright blazar 1156+295 (4C~+29.45, Ton~599) obtained with the 4.3~m Lowell Discovery Telescope. 
    During a multi-wavelength outburst in late 2017, when the $\gamma$-ray flux increased to $2.5\times 10^{-6} \; \rm phot\; cm^{-2}\; s^{-1}$ and the quasar was first detected at energies $\geq100$ GeV, the flux of the \ion{Mg}{2} $\lambda 2798$ emission line changed, as did that of the Fe emission complex at shorter wavelengths. These emission line fluxes increased along with the highly polarized optical continuum flux, which is presumably synchrotron radiation from the relativistic jet, with a relative time delay of $\lesssim2$ weeks. 
    This implies that the line-emitting clouds lie near the jet, which points almost directly toward the line of sight. The emission-line radiation from such clouds, which are located outside the canonical accretion-disk related broad-line region, may be a primary source of seed photons that are up-scattered to $\gamma$-ray energies by relativistic electrons in the jet.
\end{abstract}

\section{Introduction}\label{sec:intro}

Active galactic nuclei (AGNs) constitute the most luminous class of objects in the universe. They are also quite diverse, consisting of several different sub-classifications that depend, in part, on the angle of observation relative to the symmetry axis. The basic model of an AGN includes a central supermassive black hole (SMBH), an accretion disk (AD) around the SMBH, a dusty torus (DT) in the equatorial plane, a multitude of ionized clouds emitting spectral lines (broad and narrow emission line regions, BLR and NLR, respectively), and --- in 5-10\% of the cases --- an oppositely directed pair of relativistic plasma jets propelled out of the nucleus along the rotational axis of the AD \citep[e.g.,][]{urry1995}. When an AGN with relativistic jets is oriented such that one of the jet axes lies within $\sim10\degr$ of the line of sight, the object is classified as a blazar.  
    
Blazars dominate the $\gamma$-ray sky outside the Galactic plane. Inverse Compton (IC) scattering of UV, optical, and near-infrared photons by relativistic electrons in the jet can produce the $\gamma$-rays \citep[e.g.,][]{sikora2009}, although processes involving hadrons are also possible \citep[e.g.,][]{bottcher2013}. A major question for the IC process is the source of the seed photons that are up-scattered. They could be synchrotron photons emitted by the jet \citep[synchrotron self-Compton, or SSC, process; e.g.,][]{bloom1996}, or photons emitted from sources outside the jet (external Compton, or EC, mechanisms): the AD \citep{dermer1994}, BLR \citep{sikora1994}, and/or DT \citep{blazejowski2000, sokolov2005}. The SSC integrated flux should not greatly exceed the IR-optical synchrotron flux from the same
population of electrons, as is common during outbursts in quasars \citep{sikora2009}.
    
The relative timing of flares in $\gamma$-ray light curves, and the appearance of superluminal knots in millimeter-wavelength Very Long Baseline Array (VLBA) images, suggest that the variable $\gamma$-ray emission occurs parsecs from the SMBH \citep{jorstad2016}. This is supported by detections of very high-energy (VHE, $\geq100$ GeV) $\gamma$ rays in some quasars, since such photons cannot escape from sub-parsec regions owing to high pair-production opacities encountered as they propagate through the dense field of lower-energy photons \citep{aleksic2011}. On the other hand, $\gamma$-ray production by the EC mechanism on parsec scales is problematic, since the density of seed photons from the sources listed above should be lower than needed to explain the observed $\gamma$-ray luminosities. The zero time delay between flares in the optical and $\gamma$-ray light curves observed in a number of blazars \citep[e.g., ][]{jorstad2013,larionov2017,weaver2019} suggests that the $\gamma$-ray and optical synchrotron emission regions are co-spatial. Additionally, since the AD and main BLR are located within a parsec of the central SMBH, neither can be the main source of seed photons for IC $\gamma$-rays produced on scales of several parsecs or more. Another source of seed photons on such scales appears necessary.
    
Many solutions to this seed-photon problem have been proposed, with varying complexity. For example, \citet{ackermann2014} concluded that the $\gamma$-rays in the quasar 4C~+21.35 were produced via a combination of SSC and EC mechanisms along the jet 2-3 parsecs from the SMBH, with the EC seed photons originating from the DT. However, if the DT lies near the equatorial plane, as expected in a blazar, it is unlikely that it can provide a sufficient density of seed photons to the polar jet to explain the observed $\gamma$-ray fluxes \citep[see][]{joshi2014}. A more complex proposal is that the seed photons originate from within a slower outer layer of the jet \citep{MacDonald2015}. However, \citet{nalewajko2014} found this to be insufficient to explain the observed $\gamma$-ray flux without over-predicting the X-ray emission.
On the other hand, \citet{isler2015} have argued that in some cases the jet may provide enough ionizing photons to the canonical BLR to produce $\gamma$-rays within a parsec of the black hole, although this would require a photon field too dense to avoid the effects of pair-production opacity \citep{aleksic2011}.
    
A possible solution to the above conundrum is that the source of the seed photons is a population of emission-line clouds located near the jet, but farther than $\sim1$ pc from the black hole. \citet{leon-tavares2015} have suggested that such clouds could be entrained in an outflow surrounding the parsec-scale jet. On the other hand, \citet{punsly2013} and \citet{larionov2020} have found a time-variable red wing in the \ion{Mg}{2} $\lambda2798$ broad emission line profile in the blazar 3C~279. An increase in the flux of this red wing  during a $\gamma$-ray/optical outburst of 3C~279 in late 2016 led \citet{larionov2020} to propose that the spectral feature is emitted by polar clouds that are falling toward, rather than away from, the central SMBH. 
    
While some studies of such variability of emission lines in blazars have been carried out previously \citep[e.g.,][]{leon-tavares2013,punsly2013,leon-tavares2015,isler2015}, more detailed investigations are required to explore the general and detailed characteristics of the phenomenon and its relation to $\gamma$-ray emission. As part of an effort to provide such studies, here we present and interpret observations of the optical spectrum and optical and $\gamma$-ray light curves of the quasar 1156+195 (4C~+29.45), which has a redshift $z=0.72469\pm0.00035$ \citep[][]{hewett2010} and luminosity distance $D_\ell\approx4.5$ Gpc for a Hubble constant $H_0\approx70$ km~s$^{-1}$~Mpc$^{-1}$ within the currently favored Lambda-cold-dark-matter cosmology with $\Omega_{\textrm{matter}}\approx0.3$ and $\Omega_{\textrm{Lambda}}\approx0.7$.

\source\ is extremely variable and highly polarized at optical and radio wavelengths \citep{Wills1983, Wills1992, Fan2006, Hovatta2007, SK2008}. It is a flat-spectrum radio quasar (FSRQ) with a core-jet structure on milliarcsecond (mas) scales \citep{mchardy1990}. From the width of its \ion{Mg}{2} emission line and optical continuum flux, \citet{keck2019} derived a mass of $\sim9\times10^8 \; \rm M_\odot$ for the central SMBH. The quasar was first detected as a $\gamma$-ray source in the second  Energetic Gamma Ray Experiment Telescope (EGRET) Catalog \citep{thompson1995}, and since 2008 has been monitored by the Large Area Telescope (LAT) of the {\it Fermi} Gamma Ray Space Telescope. The VLBA-BU-BLAZAR program \citep{jorstad2016} has monitored \source\ since 2007 with monthly imaging with the Very Long Baseline Array (VLBA) within a sample of 37 bright $\gamma$-ray blazars at 43 GHz. These observations, as well as VLBA monitoring at 15 GHz \citep{Lister2016}, have determined that bright knots in the jet can move with superluminal apparent speeds up to 25$c$. From the apparent motions and the timescale of the observed decline in flux of such knots, \cite{jorstad2017} derived a Doppler factor $\delta = 12\pm3$, Lorentz factor $\Gamma = 10\pm3$, and jet viewing angle $\Theta_\circ \leq 2\fdg5$ using data from 2008-13. Based on contemporaneous $\gamma$-ray and optical light curves along with VLBA monitoring, \citet{ramakrishnan2014} constrained the location of the $\sim1$ GeV $\gamma$-ray emission in \source\ to lie close to the parsec-scale millimeter-wave core at 43~GHz. In addition, \source\ was detected for the first time at very high $\gamma$-ray energies (VHE, $\geq100$ GeV) in December 2017 by the Very Energetic Radiation Imaging Telescope Array System (VERITAS) \citep{mukherjee2017} and the Major Atmospheric Gamma Imaging Cherenkov Telescopes (MAGIC) \citep{mirzoyan2017}. 
    
In this paper, we study variability of the \ion{Mg}{2} $\lambda2798$ and blended Fe II emission lines with respect to the continuum optical and $\gamma$-ray flux of \source\ in an effort to determine the location and kinematics of the line-emitting clouds. In \S\ref{sec:obs} we present our observations of \source\ along with $\gamma$-ray and optical light curves, while in \S\ref{sec:analysis} we describe the data analysis. We discuss the results in \S\ref{sec:discuss} and draw conclusions in \S\ref{sec:conclude}. 

\section{Observations} \label{sec:obs}

\subsection{Light Curves}

We have obtained optical photometric and polarimetric measurements of \source\ in R band from \RObsStart\ to \RObsEnd, and calculated its $\gamma$-ray light curve during the same period (and through \ObsGammaEnd) using the photon and spacecraft data provided by the {\it Fermi} Science Support Center. The optical photometric and polarization data were obtained at multiple telescopes:
1) Lowell Observatory's 4.3~m Lowell Discovery Telescope (LDT; Happy Jack, AZ; formerly the Discovery Channel Telescope); 
2) 1.8~m Perkins telescope (PTO; Flagstaff, AZ);
3) 70~cm AZT-8 (Crimean Astrophysical Observatory);
4) 40~cm LX-200 (St. Petersburg, Russia);
5) 2.3~m Bok and 1.54~m Kuiper telescopes of Steward Observatory (Tuscon, AZ)\footnote{publicly available at \url{http://james.as.arizona.edu/~psmith/Fermi/}} \citep{smith2009};
and 6) 2.2~m Calar Alto telescope (Andaluc\'ia, Spain)\footnote{MAPCAT program: \url{https://www.iaa.csic.es/~iagudo/research/MAPCAT/}}.

We have used differential aperture photometry to obtain photometric measurements with comparison stars 2, 3, and 4 from \cite{Raiteri1998}, except for the Steward observatory data \citep [see][]{Smith2016}. We have obtained \NRmag\  measurements in R band, with brightness ranging from \MinR$^{\text m}$ to \MaxR$^{\text m}$. The average uncertainty of an individual measurement is $\le$0.05$^{\text m}$. Conversion of R band magnitude to flux density, $F_{\text R}$, was based on the calibration given in  \cite{Mead1990}. In order to separate optical activity states, we used the average flux density, $\langle F_{\text R}\rangle$=0.58~mJy, and its standard deviation, $\sigma_{\text R}$=0.75~mJy, reported in \cite{williamson2014}, where these values were calculated over $\sim$10 yr. We classify the flux state of the source as ``quiescent'' if $F_{\text R}\le\langle F_{\text R}\rangle$, as ``medium'' if $\langle F_{\text R}\rangle<F_{\text R}\le\langle F_{\text R}\rangle+2\sigma_{\text R}$, and as ``high'' if $F_{\text R}>\langle F_{\text R}\rangle+2\sigma_{\text R}$. This implies that a high state corresponds to R$<15.4^{\text{m}}$. 

The polarization data were obtained in R band at telescopes 2, 3, and 6 indicated above. The LX-200 telescope performed polarization measurements without a filter with central wavelength $\lambda_{\text eff}\sim 6700$ \AA.~ The Steward observatory telescopes used the Imaging/Spectro-polarimeter \citep[SPOL;][]{Schmidt1992}, which yields spectra in Stokes $q$ and $u$ parameters over a wavelength range from 4000 \AA~ to 7500 \AA~ \citep{Smith2016}. These polarization parameters are close to R-band values. A brief discussion of data reduction for all telescopes can be found in \cite{Jorstad2010}, who used the same telescopes and instruments to study the polarization behavior of the quasar 3C~454.3. The polarization parameters have been corrected for statistical bias \citep{WK1974}.

We have reduced the $\gamma$-ray data using version v1.0.10 of the {\it Fermi} Science Tools, with background models from the iso\_P8R3\_SOURCE\_V2\_v1.txt isotropic template, and the gll\_iem\_v07 Galactic diffuse emission model\footnote{\url{https://fermi.gsfc.nasa.gov/ssc/data/access/lat/BackgroundModels.html}}. We have employed an unbinned likelihood analysis of the photon data over an energy range of 0.1-200 GeV. The $\gamma$-ray emission of point sources within a $20\degr$ radius of the quasar was represented by spectral models listed in the fourth  Fermi Gamma-ray LAT catalogue (4FGL) of sources detected by the LAT \citep{Fermi2019}. The quasar's $\gamma$-ray spectrum was modeled by a log-parabola of the form
\begin{equation}
    \frac{\text{d}N}{\text{d}E} = N_0 \left( \frac{E}{E_{\text{b}}} \right)^{-(\alpha + \beta\log{E/E_{\text{b}}})}\ ,
\end{equation}
\noindent where $\alpha = 2.17$, $\beta = 6.75 \times 10^{-2}$, and break energy $E_{\text{b}}=485$ MeV, as given in the 4FGL catalogue. The spectral parameters of \source\ and all cataloged sources were kept fixed, with only the prefactor $N_0$ allowed to vary for \source, catalog sources within $5\degr$, and bright ($F_\gamma > 10^{-11}$ erg~cm$^{-2}$~s$^{-1}$) sources within $10\degr$. We binned photons over 7 days during quiescent states as defined above based on optical observations, over 4 days during a medium state, and one day during an active state. The source was considered detected if the test statistic (produced in the maximum-likelihood analysis) $TS\geq10$, which corresponds to a $\sim3\sigma$ detection level \citep{Nolan2012}. If $TS<10$, $2\sigma$ upper limits were calculated using the Python script provided by the {\it Fermi} Science Tools. The analysis resulted in \NGamma\ $\gamma$-ray measurements, out of which \NGammaUpper\ are upper limits. During the high activity state in 2017 November-December, the $\gamma$-ray flux reached a value of $(2.7\pm0.2)\times 10^{-6}$~phot~cm$^{-2}$~s$^{-1}$, which corresponds to an apparent luminosity of $(4.7\pm0.4)\times10^{48}$ erg~s$^{-1}$.

\begin{figure*}[ht]
    \plotone{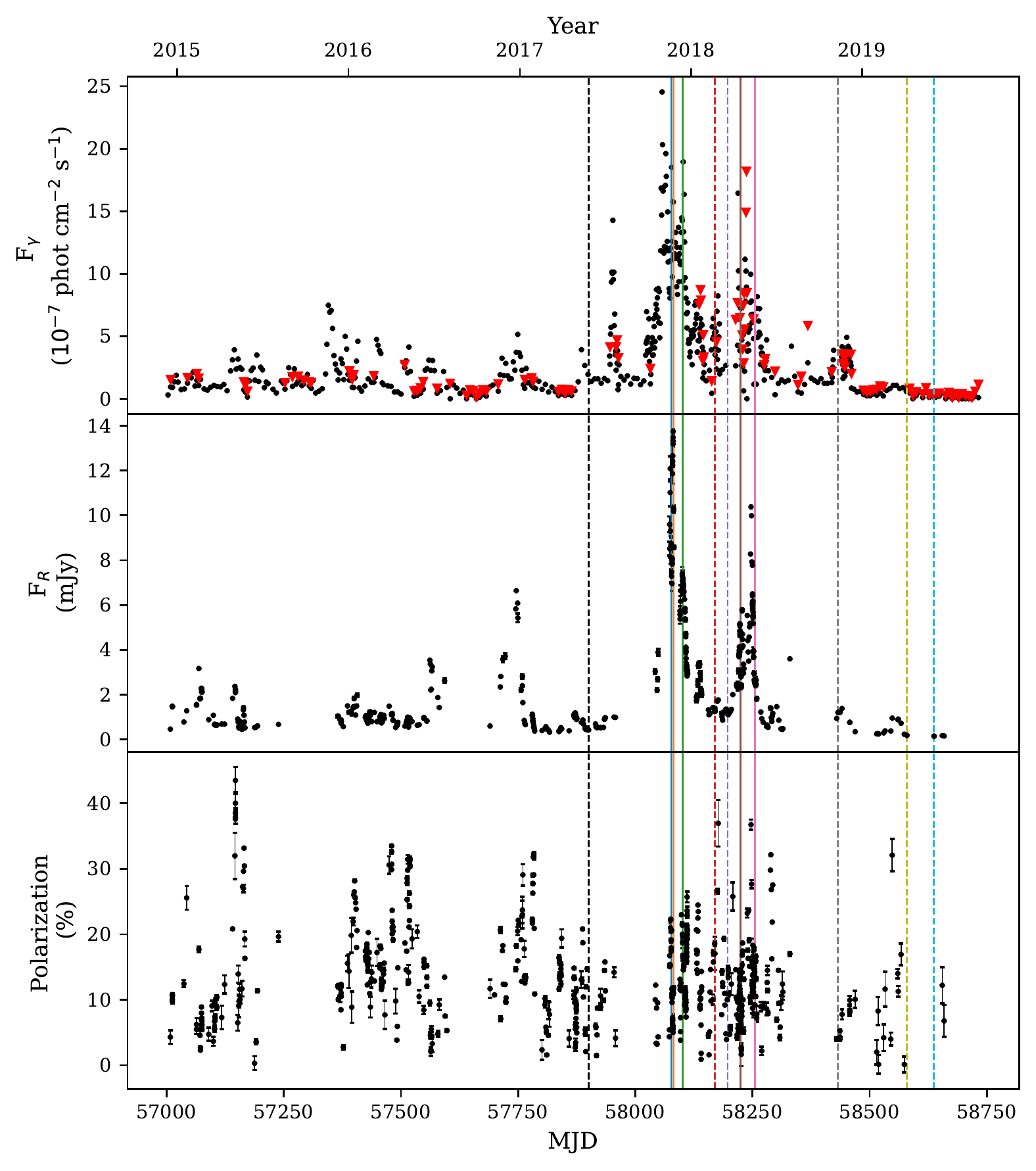}
    \caption{Gamma-ray and optical R-band light curves and polarization curve of \source\ from 2014 December to 2019 June. The colored vertical lines correspond to dates of our spectroscopic observations (see Figure~\ref{fig03}), with the dashed lines indicating a quiescent or medium state, and the solid lines indicating an active state. The red downward triangles in the top panel represent upper limits. The uncertainty in the $\gamma$-ray flux depends on the brightness of the quasar and has an average value of $1.12\times10^{-7}$ phot~cm$^{-2}$~s$^{-1}$. \label{fig01}}
\end{figure*}
    
\begin{figure}
    \centering
    \epsscale{0.5}
    \plotone{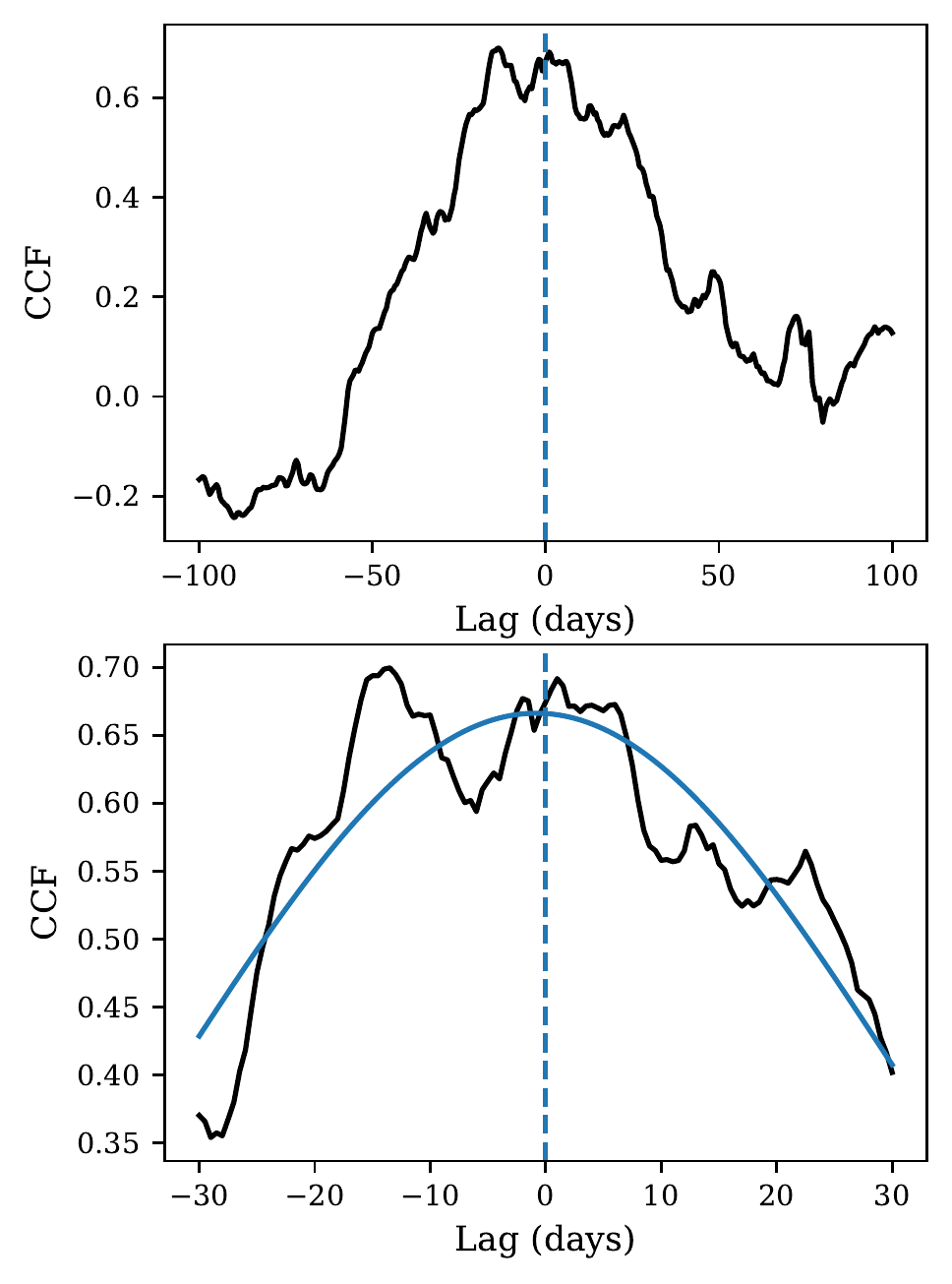}
    \caption{\textit{Top}: Cross-correlation function (CCF) between the optical and $\gamma$-ray light curves from 2016 April 22 to 2019 January 17; negative delay corresponds to the $\gamma$-ray light curve leading. \textit{Bottom}: Best Gaussian fit to the peak of the CCF, centered on $-0.81\pm0.44$ day, with FWHM $73\pm 2$ day. \label{fig02}}
\end{figure}

The $\gamma$-ray and optical R-band light curves of \source, along with the degree of polarization versus time,  are displayed in Figure~\ref{fig01}. As can be seen, the optical flare in late 2017 occurred at essentially the same time as a $\gamma$-ray flare. We have performed a cross-correlation analysis between the light curves using the PYCCF package \citep{sun2018, peterson1998}, which applies a linear interpolation between two unevenly sampled light curves. Setting the interpolation time step equal to 0.5 days, the analysis reveals a statistically significant correlation between the light curves, with a maximum correlation coefficient of 0.70 (see Figure~\ref{fig02}). By fitting a Gaussian to the CCF over lags of $\pm30$ days, we determine that the $\gamma$-ray variations lead the optical variations by $\sim 0.8 \pm 0.5$ days in the observer frame (see Figure~\ref{fig02}). The simultaneity of these flares within the uncertainties suggests that the emission sites of the two wavebands are co-spatial.

The fractional polarization of the quasar varied significantly during the outburst (from $\sim$1\% to $>$30\%), with an average value $\sim$15\%. This implies that synchrotron emission dominated the optical continuum.

\subsection{Spectral Observations} 
    
Optical spectroscopic and photometric observations of \source\ were carried out at the  LDT. We observed the quasar 12 times between 2017 May 27 and 2019 June 3. The spectroscopic observations were performed using the DeVeny spectrograph\footnote{\url{https://jumar.lowell.edu/confluence/pages/viewpage.action?pageId=23234141\#DCTInstrumentationCurrent\&Future-DeVeny}},
with a grating of 300~grooves per mm and a slit width of 2.5 arcseconds. This provided an observed wavelength range from about 3200 \AA\ to 7500 \AA, with a blaze wavelength of 5000 \AA\ and a resulting dispersion of about 2.17 \AA\ per pixel. 
    
To measure the spectrum of \source, we observed both the quasar and an A2 spectral-class comparison star, \compstar, which has a V-band magnitude of 9.83 and is located only 2.17\degr\ from the quasar. An observation started with $\sim$2 exposures of the comparison star, followed by $\sim$3 exposures of \source, and finished with $\sim$2 more exposures of \compstar. Additionally, we observed \source\ using the Large Monolithic Imager (LMI) at the LDT just before or after a spectral observation. We took $\sim$3 exposures in R and V bands to provide photometry for the flux density calibration of the spectrum. A summary of these observations can be found in Table~\ref{tab1}.
   
\begin{deluxetable*}{@{\extracolsep{4pt}}cccccccccccc@{}}[htb!]
    \tablecaption{LDT Observations of 1156+295 \label{tab1}}
    \tablehead{
        \colhead{} & \multicolumn{3}{c}{DeVeny} & \multicolumn{6}{c}{LMI} & \colhead{} \\ 
        \cline{2-4} \cline{5-10}
        \colhead{} & \multicolumn{3}{c}{} & \multicolumn{3}{c}{V} & \multicolumn{3}{c}{R} & \colhead{} \\
        \cline{5-7} \cline{8-10}
        \colhead{Date} & \colhead{MJD} & \colhead{N} & \colhead{Exp.} & \colhead{N} & \colhead{Exp.} & \colhead{mag} & \colhead{N} & \colhead{Exp.} & \colhead{mag} & \colhead{State} \\ 
        \colhead{(yyyy-mm-dd)} & \colhead{} & \colhead{} & \colhead{(s)} & \colhead{} & \colhead{(s)} & \colhead{} & \colhead{} & \colhead{(s)} & \colhead{} & \colhead{} \\ 
        \colhead{(1)} & \colhead{(2)} & \colhead{(3)} & \colhead{(4)} & \colhead{(5)} & \colhead{(6)} & \colhead{(7)} & \colhead{(8)} & \colhead{(9)} & \colhead{(10)} & \colhead{(11)}
    }
    \startdata
        2017-05-27 & 57900.24 & 2 & 1200 & 2 &120& $17.59 \pm 0.01$ & 2 & 90 & $17.19 \pm 0.01$ & Q \\
        2017-11-20 & 58077.49 & 3 & 600 & 0 & ... & ... & 3 & 30 & $14.08 \pm 0.02$ & A \\
        2017-11-25 & 58082.48 & 3 & 900 & 2 & 10 & $14.40 \pm 0.01$ & 2 & 10 & $13.93 \pm 0.01$ & A\\
        2017-12-14 & 58101.49 & 3 & 900 & 2 & 15 & $14.66 \pm 0.01$ & 2 & 10 & $14.23 \pm 0.01$ & A\\
        2017-12-24\tablenotemark{a} & 58111.40 & 3 & 900 & 0 & ... & ... & 3 & 60 & $15.10 \pm 0.01$ & A \\
        2018-02-21 & 58170.45 & 2 & 1200 & 2 & 30 & $16.47 \pm 0.01$ & 2 & 30 & $16.01 \pm 0.01$ & M \\
        2018-03-20 & 58197.45 & 2 & 900 & 2 & 15 & $16.40 \pm 0.01$ & 2 & 10 & $15.98 \pm 0.01 $ & M \\
        2018-04-16  & 58224.15 & 5 & 750 & 2 & 15 & $15.07 \pm 0.01$ & 2 & 10 & $14.59 \pm 0.01$  & A\\
        2018-05-17 & 58255.26 & 3 & 1200 & 2 & 25 & $15.71 \pm 0.01$ & 2 & 15 & $15.25 \pm 0.01$ & A \\
        2018-11-10 & 58432.50 & 3 & 900, 600 & 2 & 60 & $16.50 \pm 0.01$ & 2 & 45 & $16.06 \pm 0.01$ & M \\
        2019-04-06 & 58579.30 & 3 & 1500 & 0 & ... & ... & 3 & 15 & $18.12 \pm 0.01$ & Q \\
        2019-06-03 & 58637.23 & 3 & 1500 & 3 & 40 & $18.60 \pm 0.01$ &3 & 25 & $18.35 \pm 0.01$ & Q \\
    \enddata
    \tablecomments{Columns include: (1) The UT date of observations; (2) The UT of the DeVeny observations; (3, 5, 8) The number of exposures; (4, 6, 9) The exposure times in seconds; (7, 10) The measured magnitude of the quasar; (11) The state of the quasar during observations (``A'' for active, ``M'' for medium, and ``Q'' for quiescent). \\
    \tablenotemark{a} Spectra from this date were excluded from the analysis due to bad weather conditions.}
\end{deluxetable*}

Each spectral image of the quasar and comparison star was checked for systematic errors, cleaned for cosmic rays, and corrected for bias and flat-field effects. 
The corrected spectra of \compstar, obtained before and after the quasar exposures, were summed. The same was done for the source spectra. This provided a signal-to-noise ratio for the  \source\ and \compstar\ spectra $\ge$25 and $>$100, respectively.

We followed the technique of spectral reduction described in \cite{vacca2003} and realized in an IDL program (v.8.6). Since \compstar\ has a spectral class of A2, we have assumed that it has about the same intrinsic spectrum as Vega (A0). We performed shifting, scaling, and reddening of the Vega model spectrum \citep{Vega1994} to match the summed spectrum of the comparison star. This allowed us to determine the point spread function (PSF), atmospheric extinction, and position and profile of telluric lines during our observations, which were applied to calibrate the quasar's spectrum. 
We have estimated the uncertainty of the measured flux density in each spectrum by considering the following sources of error: (1) signal-to-noise ratio of the data (SNR); (2) errors introduced during the reduction process; and (3) uncertainty in the wavelength dispersion.
The SNR was determined by using the technique described in \cite{stoehr2008}. \textbf{A total uncertainty in the flux of the emission lines of 2-5\% was then calculated as the sum of these uncertainties added in quadrature.}

\begin{figure*}[ht]
    \plotone{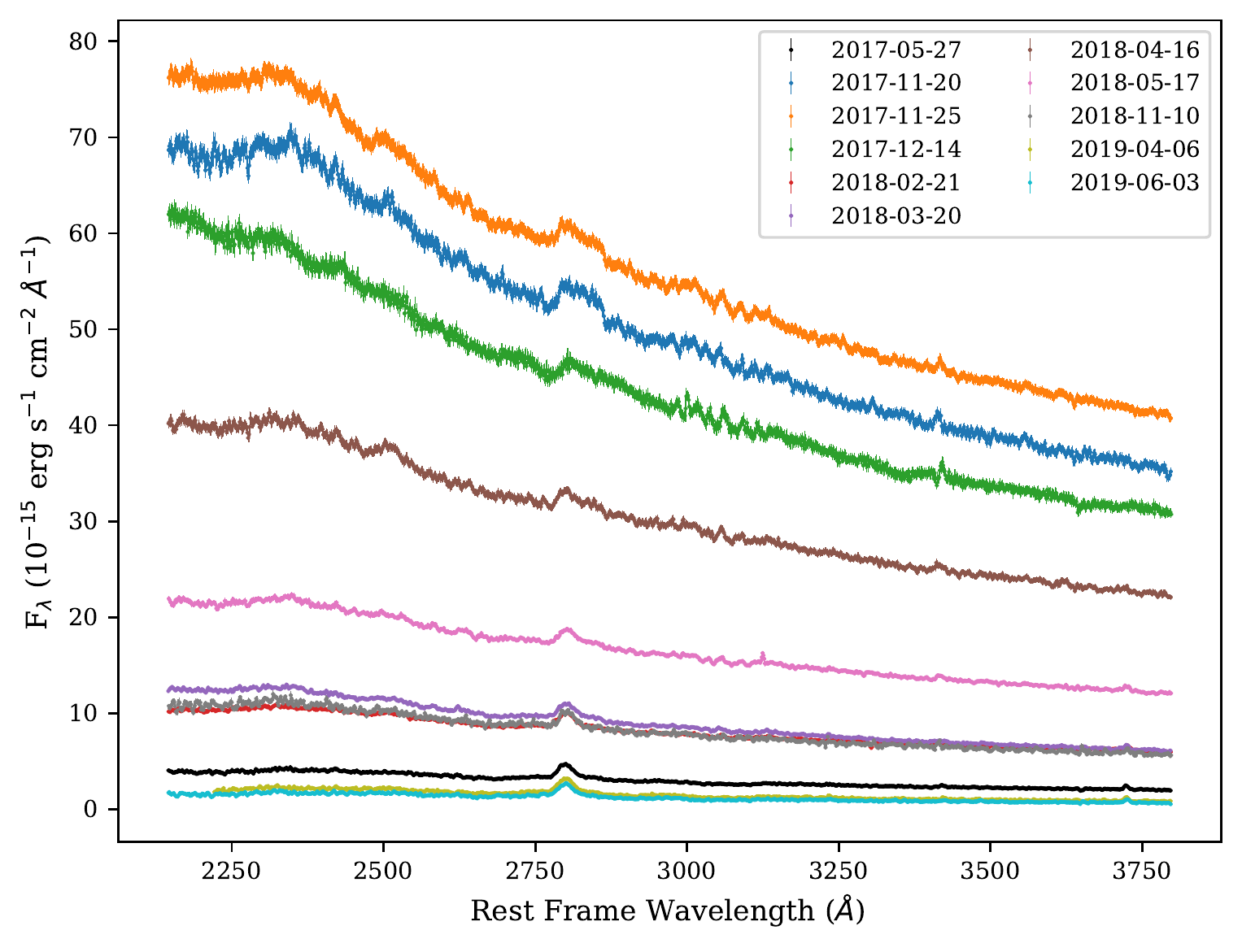}
    \caption{Optical spectra of 1156+295 \label{fig03}}
\end{figure*}

We have shifted the calibrated spectrum of the quasar into the rest frame using cosmological correction factors of $(1+z)^{-1}$ and $(1+z)^3$ for the wavelength $\lambda$ and flux density $F_\lambda$, respectively. The resulting reduced spectra are displayed in Figure~\ref{fig03}.

\section{Spectral Analysis}\label{sec:analysis}

The spectra presented in Figure~\ref{fig03} contain a prominent \ion{Mg}{2} emission line at a rest-frame wavelength of 2798 \AA, whose red wing rises as the continuum flux increases, as well as blended \ion{Fe}{2} emission lines on both sides of \ion{Mg}{2}, which also intensify with the continuum. To obtain parameters of the emission lines as a function of the continuum flux, we have modeled each spectrum as described below. 

\subsection{Continuum Modeling}\label{sec:cont_rem}

The optical continuum emission from a blazar-class quasar is often modeled as a combination of thermal radiation from an accretion disk and synchrotron emission from a relativistic jet \citep[e.g.,][]{isler2015, keck2019}. Since the synchrotron dominates over the thermal emission in 1156+295 \citep{ramakrishnan2014}, as is the case for many blazars \citep[e.g.,][]{ghisellini2014, larionov2020}, the continuum can be modeled as a simple power law:
\begin{equation}\label{eq:cont}
    F_\lambda = F_\lambda(3000\;\text{\AA}) \left(\frac{\lambda}{3000\; \text{\AA}}\right)^{\alpha},
\end{equation}
where F$_\lambda$(3000\AA) is the flux density at the arbitrarily chosen reference wavelength of 3000 \AA,  and $\alpha$ is the spectral index.
A possible complicating factor is the presence of the Balmer continuum, as described in \citep{kovacevic2014, grandi1982}. However,
in the case of 1156+295, the synchrotron radiation is much brighter than the Balmer continuum radiation; therefore, we model the continuum as a simple power law.

AGNs often produce broad Fe-complex emission features, perhaps resulting from micro-turbulence in a confined gas (such as winds) or collisional excitation in warm, dense gas \citep{baldwin2004}. Owing to the complexity of the electron energy levels of Fe, many Fe lines form at closely spaced wavelengths. These blend together to form broad pseudo-continuum emission features owing to a range of Doppler shifts from motions of the many clouds that emit the lines. The Fe emission must be taken into account when modeling the continuum and \ion{Mg}{2} $\lambda$2798 spectral line. 
Therefore, we fit the power-law continuum model only in regions without significant Fe II or III emission.

\begin{deluxetable*}{c c c c c}[htb!]
    \tablecaption{Fe Windows  \label{tab2}}
    \tablehead{
        \colhead{(1)} & 
        \colhead{(2)} &
        \colhead{(3)} &
        \colhead{(4)} &
        \colhead{(5)}
    }
    \startdata
        2250-2320 & 2333-2445 & 2470-2625 & 2675-2755 & 2855-3010 \\
    \enddata
    \tablecomments{All window ranges are in \AA.}
\end{deluxetable*}

\begin{deluxetable}{c c c}[htb!]
    \tablecaption{Continuum Windows  \label{tab3}}
    \tablehead{
        \colhead{(1)} & 
        \colhead{(2)} &
        \colhead{(3)} 
    }
    \startdata
        2200-2230\tablenotemark{a} & 3030-3090\tablenotemark{b} & 3540-3600\tablenotemark{b} \\
    \enddata
    \tablecomments{All window ranges are in \AA.\\
    \tablenotemark{a} from \cite{sameshima2011} \\
    \tablenotemark{b} from \cite{tsuzuki2006}}
\end{deluxetable}

\begin{figure*}[ht]
    \plotone{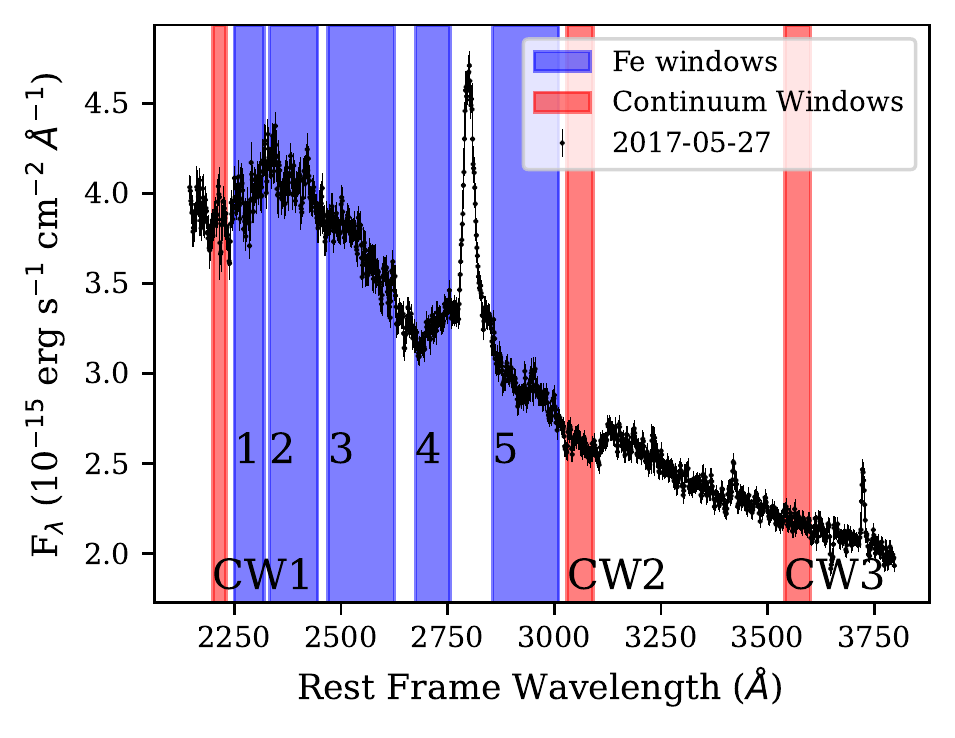}
    \caption{Spectrum of \source\ on 2017 May 27 with the Fe II/III complex and continuum windows.}
    \label{fig04}
\end{figure*}

Table~\ref{tab2} lists the Fe-complex windows present in our spectra, as identified by \citet{vestergaard2001}, while Table~\ref{tab3}
gives the continuum windows with little or no line emission, which we have used for fitting the continuum. Figure~\ref{fig04} plots the Fe and continuum windows for one of our spectra.

We carried out the power law continuum fitting over each spectrum in the indicated continuum windows using the Levenberg-Marquardt least squares method with the LMFIT python package \citep{newville2014}. The fitted parameters are given in Table~\ref{tab4}, and Figure~\ref{fig05} show the fits for all spectra.

\begin{deluxetable*}{c C C C C}[htb!]
    \tablecaption{
        Continuum Fitting Results \label{tab4}
        }
    \tablehead{
        \colhead{Epoch} &
        \colhead{SNR} &
        \colhead{$\alpha_{\lambda}$} &
        \colhead{F$_\lambda (3000)$} &
        \colhead{$\chi^2_{\textrm{dof, Fe}}$} \\
        \colhead{yyyy-mm-dd} &
        \colhead{} &
        \colhead{} &
        \colhead{$\mathrm{10^{-15}~ erg~s^{-1}~cm^{-2}} \text{\AA}^{-1}$} &
        \colhead{} \\
        \colhead{(1)} &
        \colhead{(2)} &
        \colhead{(3)} &
        \colhead{(4)} &
        \colhead{(5)}
        }
    \startdata 
        2017-05-27 & 75 & -1.207 \pm 0.009 & 2.652 \pm 0.004 & 0.689 \\
        2017-11-20 & 159 & -1.220 \pm 0.008 & 46.850 \pm 0.065 & 1.269 \\
        2017-11-25 & 333 & -1.148 \pm 0.004 & 53.061 \pm 0.042 & 0.606 \\
        2017-12-14 & 132 & -1.259 \pm 0.006 & 40.746 \pm 0.040 & 0.552 \\
        2018-02-21 & 154 & -0.985 \pm 0.006 & 7.554 \pm 0.008 & 0.713 \\
        2018-03-20 & 142 & -1.320 \pm 0.007 &  8.252 \pm 0.011 & 1.014 \\
        2018-04-16 & 173 & -1.068 \pm 0.006 & 28.555 \pm 0.028 & 0.735 \\
        2018-05-17 & 187 & -1.048 \pm 0.006 &  15.429 \pm 0.017 & 0.902 \\
        2018-11-10 & 60 & -1.197 \pm 0.012 & 7.493 \pm 0.015 & 0.936 \\
        2019-04-06 & 30 & -1.590 \pm 0.040 &  1.202 \pm 0.006 & 1.245 \\
        2019-06-03 & 27 & -1.517 \pm 0.032 & 0.947 \pm 0.005 & 1.387
    \enddata
    \tablecomments{Columns are: (1) UT date of observation; (2) estimated SNR of the spectrum; (3) spectral index; (4) continuum flux density at 3000 \AA;  (5) reduced $\chi^2$ value for the \ion{Fe}{2} template fit. 
    }
\end{deluxetable*}

\begin{figure*}[!bp]
    \centering
    {\plottwo{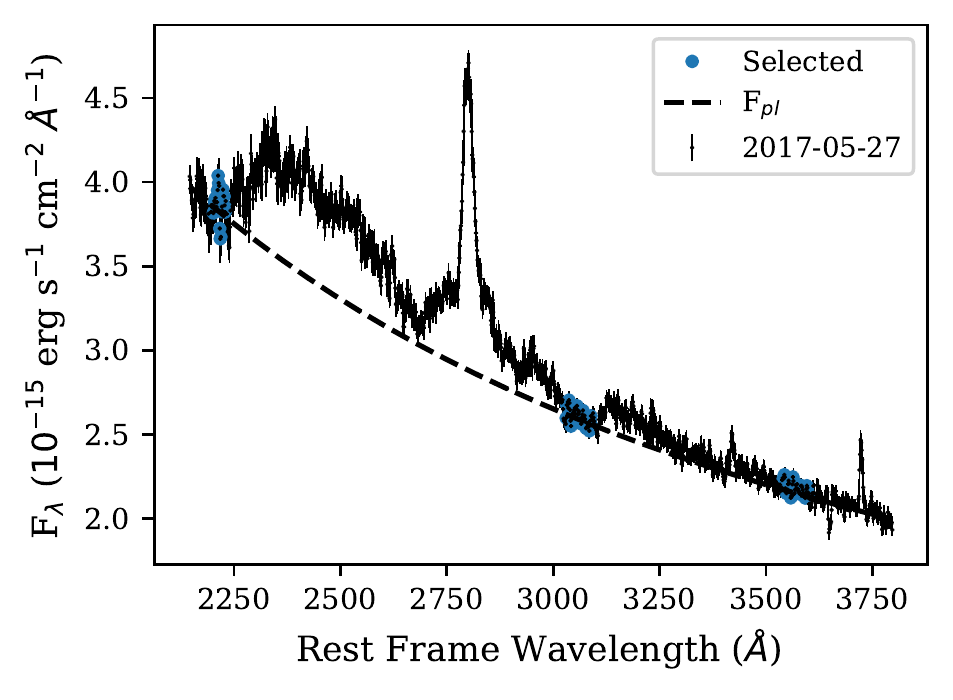}{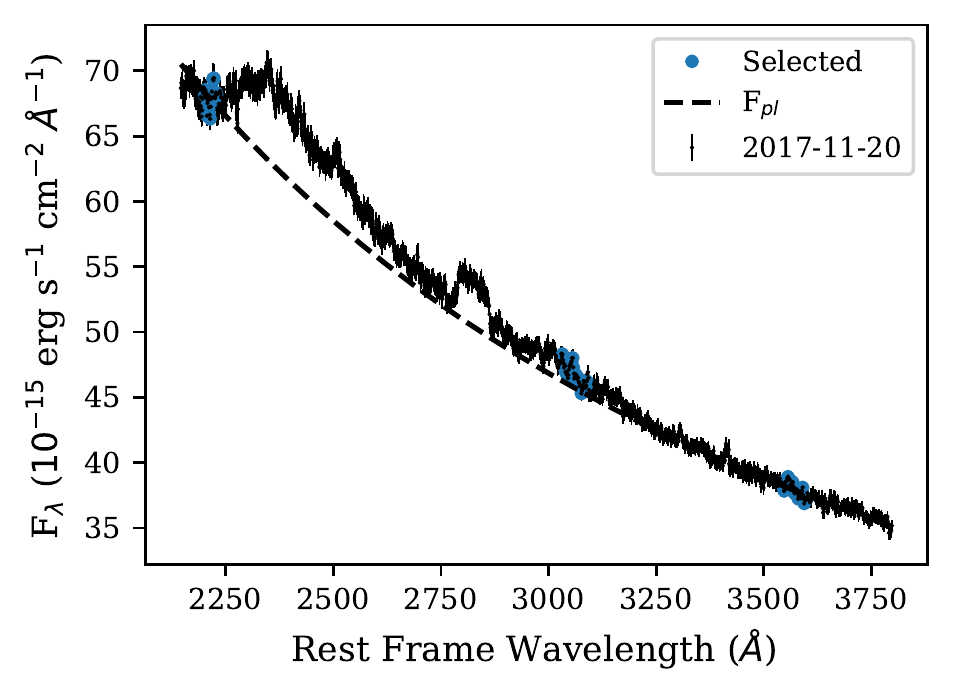}}\\
    {\plottwo{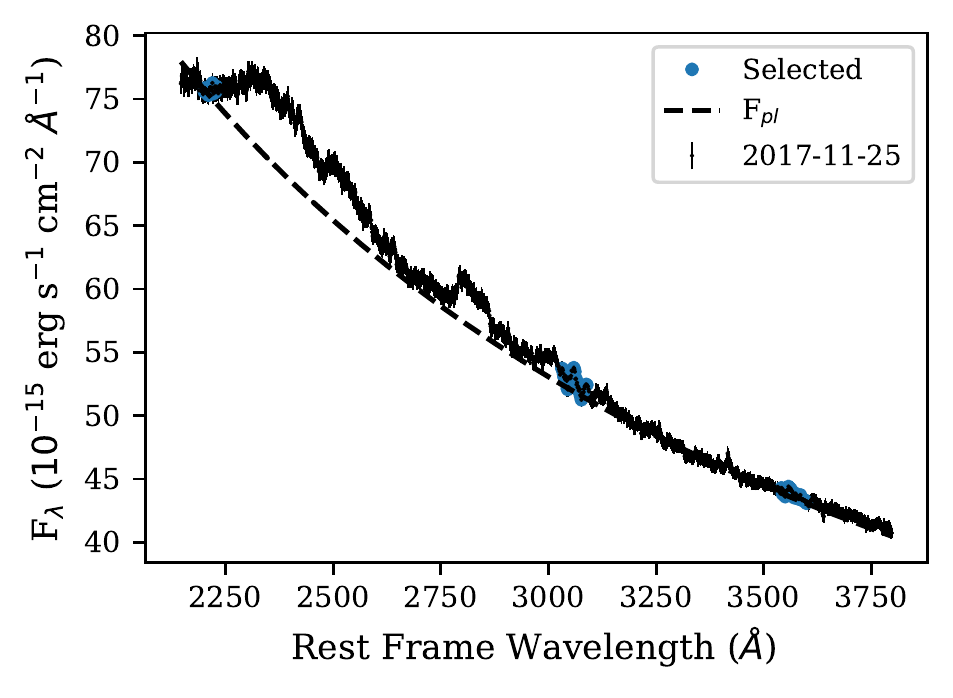}{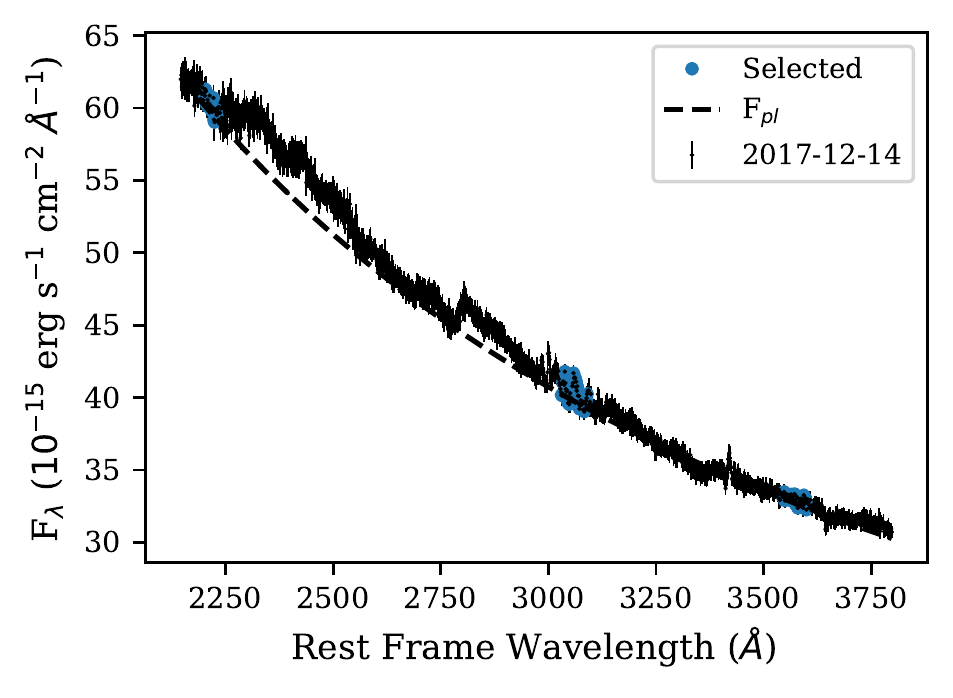}}\\
    {\plottwo{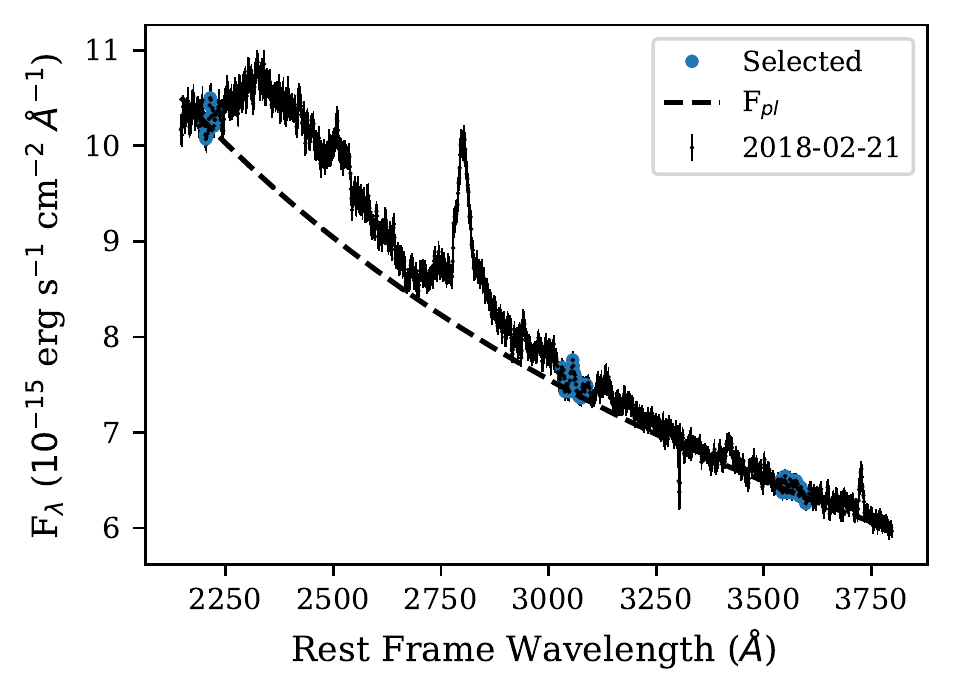}{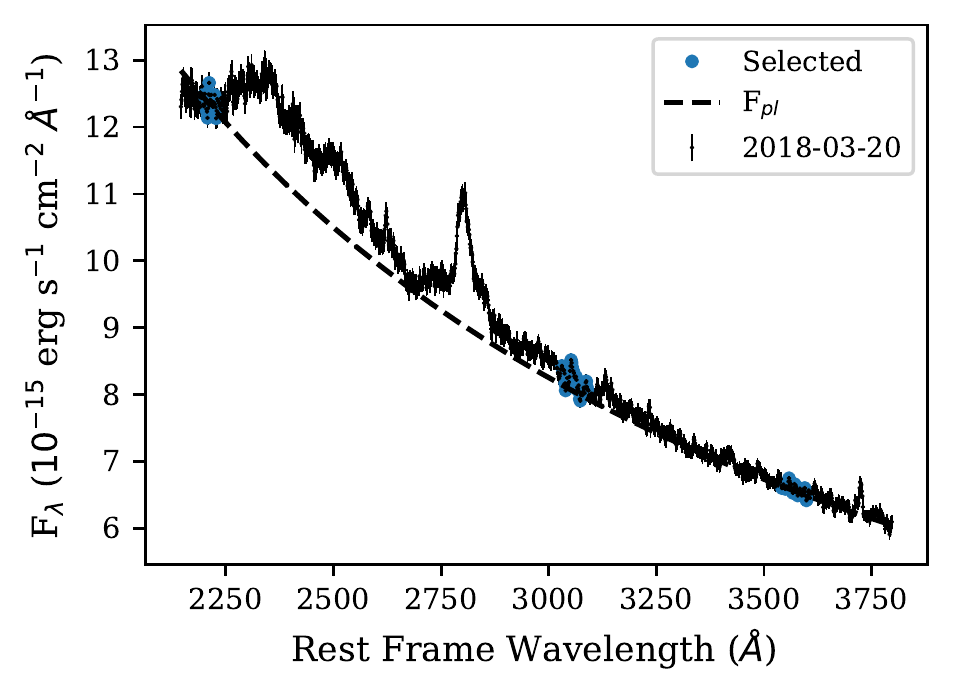}}
    \caption{Spectra plotted with the best-fit continuum (dashed line).}
\end{figure*}

\begin{figure*}[!tbp]
    \ContinuedFloat
    \centering
    {\plottwo{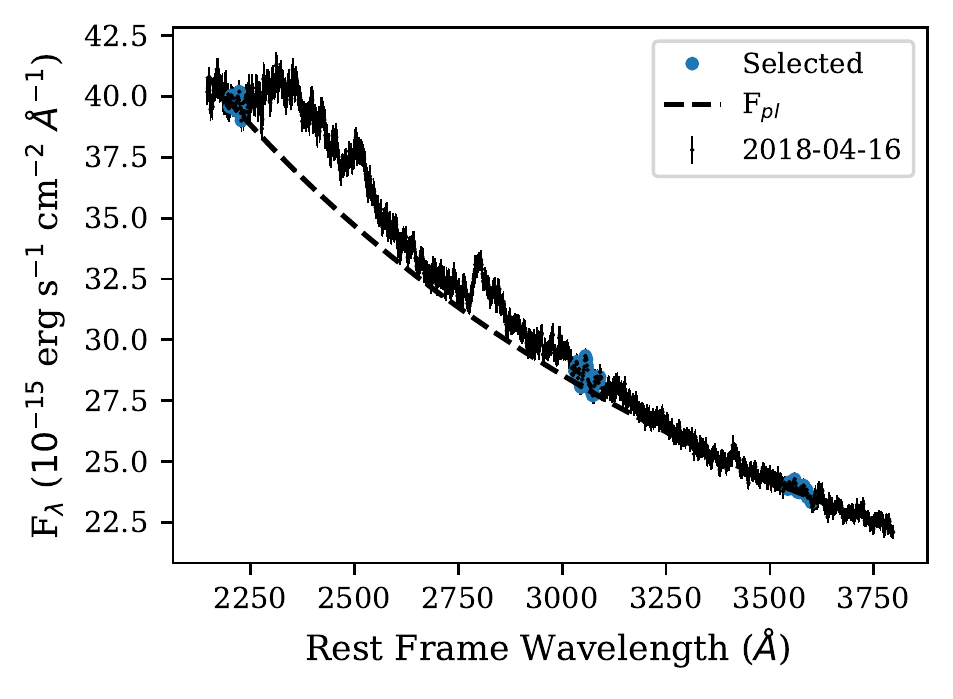}{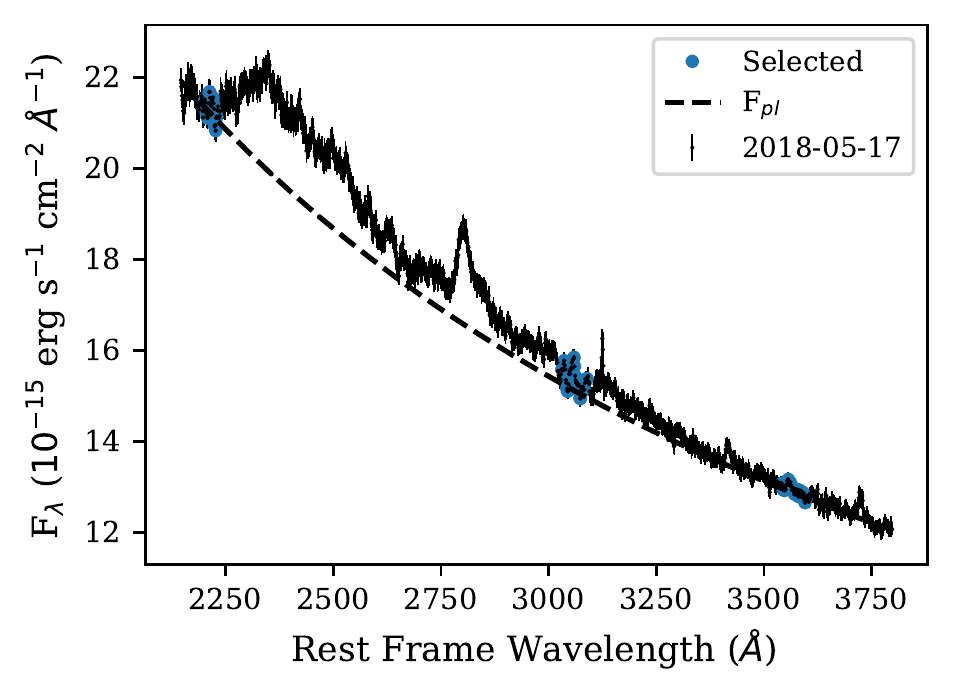}}\\
    {\plottwo{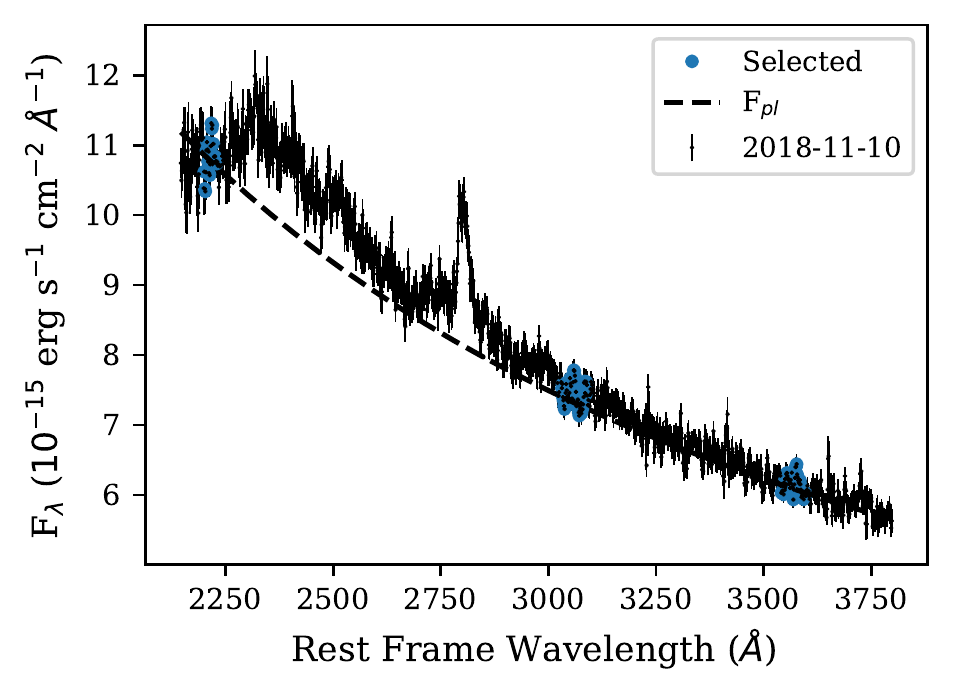}{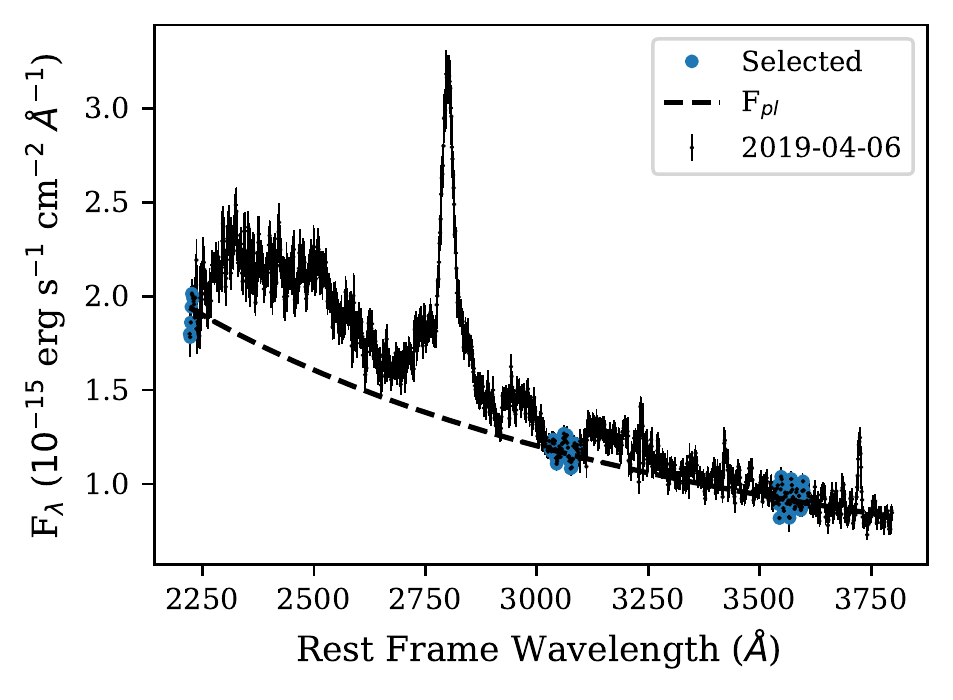}}\\
    \epsscale{0.5}
    {\plotone{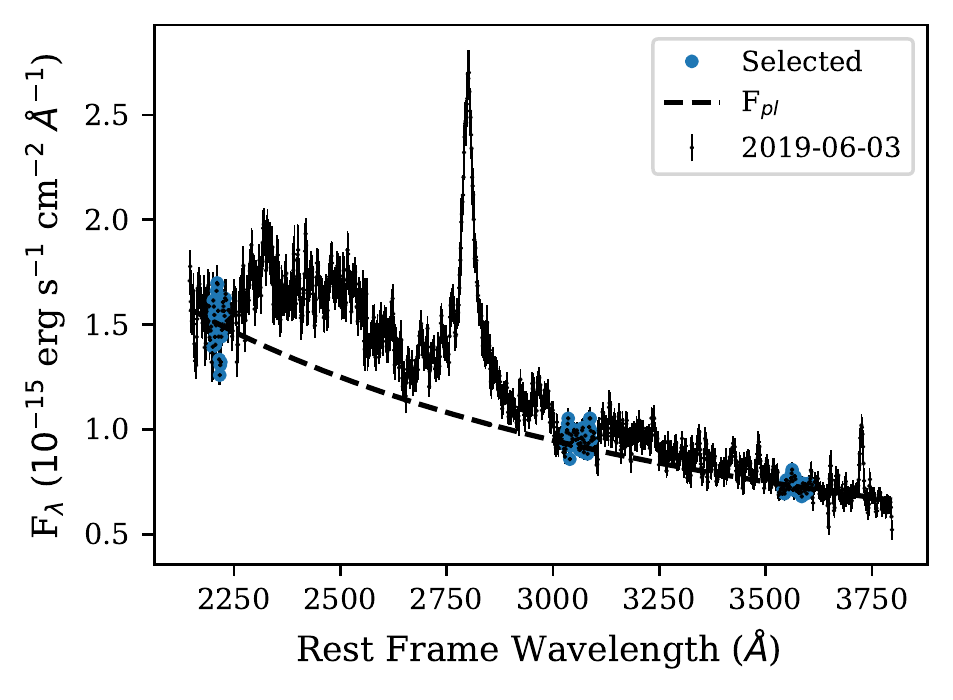}}
    \caption{Continued.}
    \label{fig05}
\end{figure*}

\subsection{Mg II and Fe-complex Modeling}

The \ion{Mg}{2} $\lambda2798$ line is the most prominent emission line in our spectra. As can be seen in Figure \ref{fig03}, there appears to be a component redward of the core line centered at 2798 {\AA}. This red wing appears to become brighter at higher flux states. We therefore model the \ion{Mg}{2} $\lambda2798$ emission line profile as the sum of two Gaussians (a ``core`` and ``wing`` component), each defined by:
\begin{equation}
    f(\lambda;A,\mu,\sigma) = \frac{A}{\sigma\sqrt{2\pi}} \exp{\left(-\frac{(\lambda-\mu)^2}{2\sigma^2}\right)}, 
\end{equation} 
where $A$ is the amplitude, $\mu$ is the central wavelength of the emission line, and $\sigma$ is the standard deviation. The full-width at half-maximum (FWHM) is then defined as:
\begin{equation}
    \sigma_{\rm{FWHM}} = 2\sqrt{2\ln{(2)}}\sigma,
\end{equation}
and is converted from wavelength space to velocity space using
\begin{equation}
    \Delta v_{\rm{FWHM}} = c \frac{\sigma_{\rm{FWHM}}}{\mu},
\end{equation}
where $c$ is the speed of light.

As stated previously, our spectra consist of blended Fe II and III line complexes. Because these complexes extend to the region of the \ion{Mg}{2} $\lambda 2798$ emission line, it can be difficult to separate the \ion{Mg}{2} lines from the Fe lines. To aid in doing so, we have utilized the \ion{Fe}{2} and \ion{Fe}{3} emission-line template composed by \cite{vestergaard2001}. The Fe template was first convolved (in logarithmic wavelength space) with a Gaussian function that broadened the template lines to 3000 km s$^{-1}$ (FWHM), as described by \cite{vestergaard2001}. This FWHM was adopted because it is close to the average FWHM of our observed \ion{Mg}{2} emission lines. (While in some cases one could attempt to determine the broadening using individual Fe lines in some regions of the spectra, e.g., \citealt{boroson1992}, this is not possible for our observed spectra owing to the blended nature of the Fe lines in this region). 

The IDL MPFIT fitting routine (which uses Levenberg-Marquardt least-squares minimization), as implemented by the python package hyperspy \citep{hyperspy}, was used to fit the double Gaussian profile and the broadened Fe template, with a scaling factor applied to each continuum-subtracted spectrum. We performed the fit in the 2650-2950 \AA\ range of the rest-frame spectra. This region was chosen in order to include the \ion{Mg}{2} emission line, as well as Fe windows 4 and 5 (as indicated in Table \ref{tab2}). We note that the Fe-complex template provided by \cite{vestergaard2001} contains multiple sub-templates. Of these, we use the UVA template and the UV47 template. The UV47 template is within the wavelength range of our spectra, but not within the range over which we perform our fitting routine. Because of this, we assume that UVA and UV47 have the same scaling factor. The parameters of the fit are reported in Table~\ref{tab5}, and the fits are plotted in Figure~\ref{fig06}. Additionally, we note that Table~\ref{tab5} reports the shift of the center of each of the fitted Gaussian components with respect to 2798 {\AA}, a positive shift indicates that the Gaussian is centered redward of 2798 {\AA}.

According to Table~\ref{tab5} the average shift of the central wavelength relative to that corresponding to the published redshift is only $123 \pm 128$ and $5095 \pm 2932$ km s$^{-1}$ for the core and wing components, respectively, of the \ion{Mg}{2} $\lambda 2798$ emission line in the quiescent state, but $660 \pm 208$ and  $6059 \pm 1488$ km s$^{-1}$ in the active state. Thus, there appears to be no significant shift of the core component during quiescent states, but there may be a small shift redward during active states. Given the error ranges of the shifts of the wing component during active and quiescent states, it appears that the center of the wing does not change significantly from quiescent to active states, but remains several thousand km s$^{-1}$ redward of the line core. Table~\ref{tab5} also shows that the red wing does generally increase in brightness during higher flux states.

As can be seen in Figure~\ref{fig06}, the Fe template fits the emission well around the \ion{Mg}{2} $\lambda2798$ line, but the template underestimates the Fe emission in the broad feature blueward of the \ion{Mg}{2} line. This is discussed in greater detail in the following section.

\begin{longrotatetable}
\begin{deluxetable}{@{\extracolsep{4pt}} c C C C C C C C C C C@{}}
    \tablecolumns{11}
    \tablecaption{Results of \ion{Mg}{2} $\lambda2798$ Emission Line Fitting \label{tab5}}
    \tablehead{
        \colhead{Epoch} &
        \multicolumn{2}{c}{Shift} &
        \multicolumn{2}{c}{FWHM} &
        \multicolumn{2}{c}{Flux} &
        \colhead{UVA} &
        \colhead{$\chi^2_{\rm dof}$}
        \\ 
        \colhead{yyyy-mm-dd} & 
        \multicolumn{2}{c}{$\mathrm{km\,s^{-1}}$} & 
        \multicolumn{2}{c}{$\mathrm{km\,s^{-1}}$} & 
        \multicolumn{2}{c}{$\mathrm{10^{-15} erg\,s^{-1}\,cm^{-2}}$} & 
        \colhead{}&
        \colhead{} 
        \\
        \cline{2-3}
        \cline{4-5}
        \cline{6-7}
        &
        \colhead{Core} &
        \colhead{Wing} &
        \colhead{Core} & 
        \colhead{Wing} &  
        \colhead{Core} &
        \colhead{Wing} &
        \colhead{} &
        \colhead{}
        \\
        \colhead{(1)} &
        \colhead{(2)} &
        \colhead{(3)} &
        \colhead{(4)} &
        \colhead{(5)} &
        \colhead{(6)} &
        \colhead{(7)} &
        \colhead{(8)} &
        \colhead{(9)}
        \\
        }
    \startdata
        2017-05-27 & 146 \pm  30 & 5305 \pm 238 & 
        3948 \pm  62 &  1753 \pm 582 &
         67.2 \pm  1.0 & 1.9 \pm  0.6 & 
        0.051 \pm 0.001 & 1.33 
        \\ 
        2017-11-20 & 1136 \pm  93 & 5700 \pm 147 &
        4947 \pm 247 & 1786 \pm 354 &
        185.1 \pm  7.4 & 25.2 \pm  5.4 &
        0.232 \pm 0.009 & 0.58
        \\ 
        2017-11-25 & 736 \pm  64 & 5787 \pm 115 &
        5035 \pm 171 & 2104 \pm 273 &
        176.6 \pm  4.9 & 26.4 \pm  3.6 &
        0.192 \pm 0.006 & 0.27
        \\
        2017-12-14 & 1237 \pm  89 & 8933 \pm 260 & 
        3817 \pm 208 & 6954 \pm 727 &
        84.1 \pm  4.4 & 71.5 \pm  7.5 &
         0.144 \pm 0.007 & 0.26
        \\
        2018-02-21 & 275 \pm  31 & 5587 \pm 269 &
         4194 \pm  77 & 2307 \pm 660 &
         79.3 \pm  1.2 &  3.4 \pm  0.9 &
         0.067 \pm 0.002 & 0.71
         \\
        2018-03-20 & 392 \pm  35 & 5645 \pm 141 & 
        4445 \pm  86 & 1980 \pm 345 &
        82.9 \pm  1.3 & 5.6 \pm  0.9 &
        0.058 \pm 0.002 & 0.65
        \\
        2018-04-16 & 457 \pm  66 & 5100 \pm 226 &
        4016 \pm 170 & 2428 \pm 540 &
        98.1 \pm  3.4 &  13.8 \pm  3.1 &
        0.146 \pm 0.005 & 0.46
        \\
        2018-05-17 & 456 \pm  37 & 5885 \pm 177 & 
        4252 \pm  90 & 1438 \pm 424 &
         89.0 \pm  1.6 & 3.6 \pm  1.0 &
          0.113 \pm 0.002 & 0.44
        \\
        2018-11-10 & 399 \pm  52 &  6117 \pm 133 & 
        3701 \pm 119 & 668 \pm 331 &
         74.7 \pm  2.1 & 1.9 \pm  0.8 &
          0.071 \pm 0.003 & 0.90
        \\
        2019-04-06 & 112 \pm  34 & 5321 \pm 270 & 
        4225 \pm  75 & 1389 \pm 661 &
         72.7 \pm  1.2 & 1.3 \pm  0.6 &
         0.047 \pm 0.001 & 1.33
        \\
        2019-06-03 & 96 \pm  41 & 5012 \pm 110 & 
        4327 \pm  94 & 307 \pm 237 &
         61.5 \pm  1.2 & 0.5 \pm  0.3 &
         0.041 \pm 0.001 & 1.73
        \\
    \enddata
    \tablecomments{Columns correspond to: (1) UT date of observation; (2-3) shift of the center of the \ion{Mg}{2} measured line from 2798 \AA\ rest wavelength for the core and wing component, respectively; (4-5) FWHM of line profile for the core and wing component of \ion{Mg}{2}, respectively; (6-7) flux of the core and wing components of Mg II line, respectively; (8) scale factor used for the Fe II/III template (9) reduced $\chi^2$ of the fit. 
    }
\end{deluxetable}
\end{longrotatetable}

\begin{figure*}[!bp]
    \centering
    {\plottwo{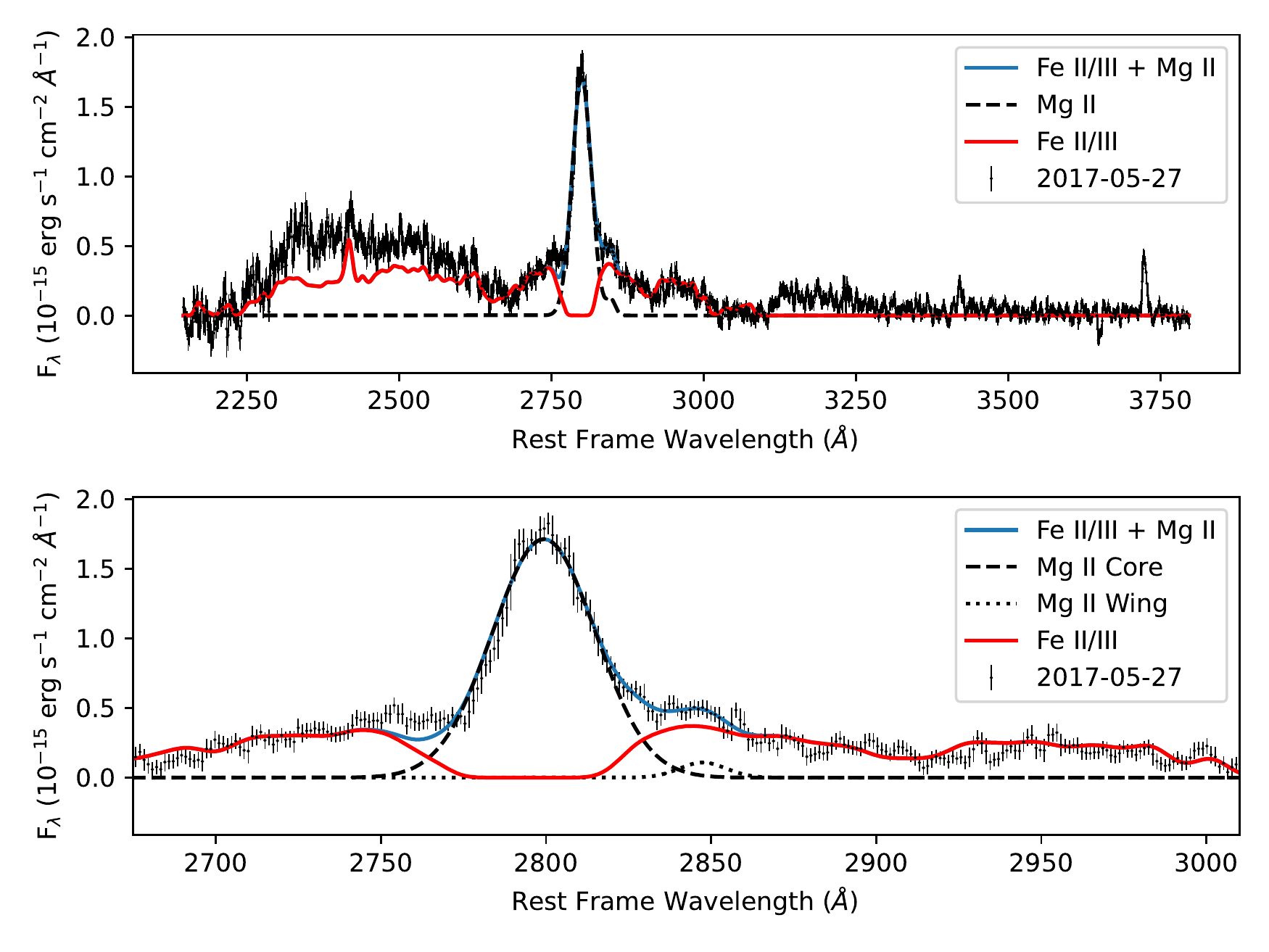}{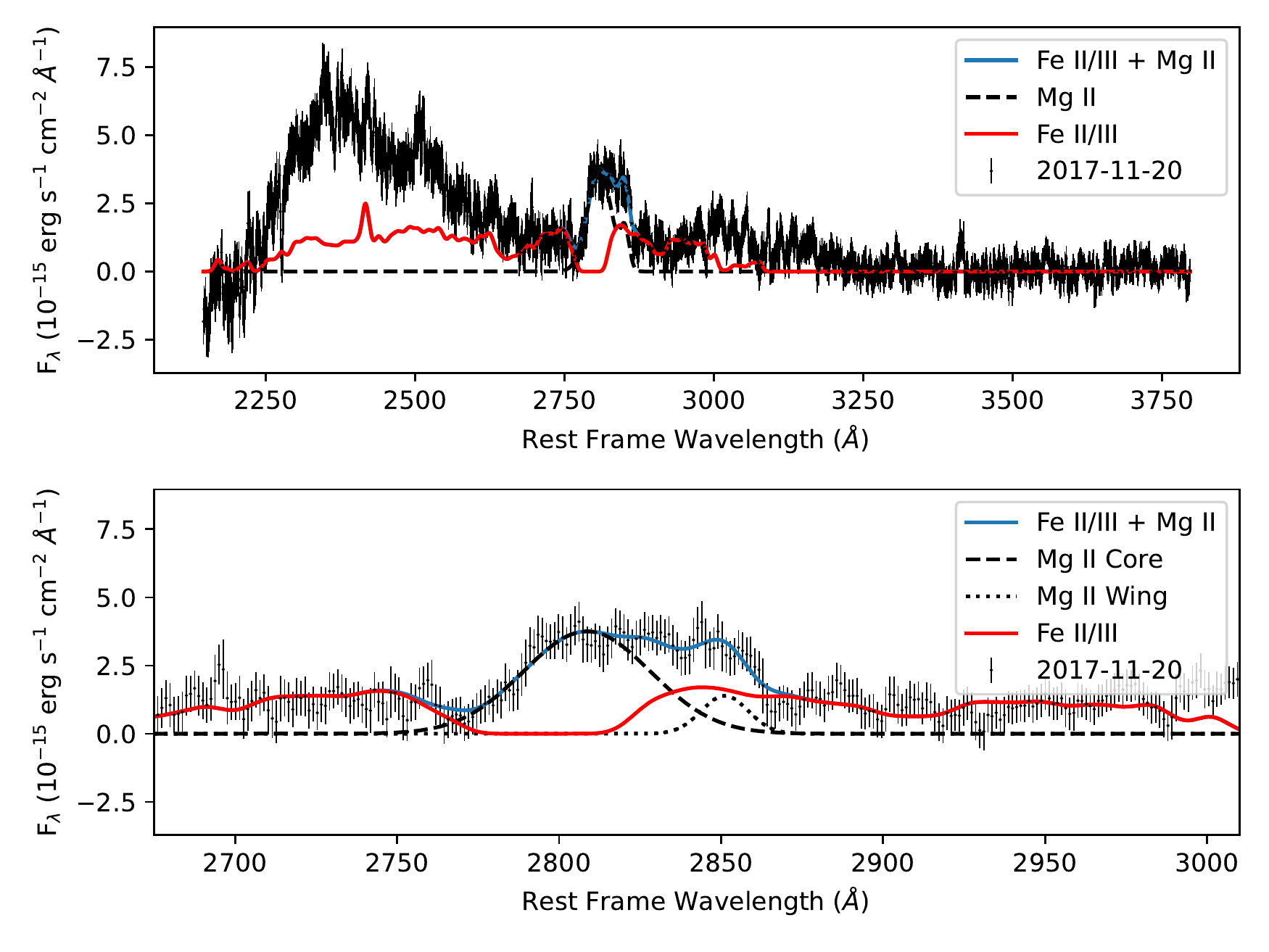}}\\
    {\plottwo{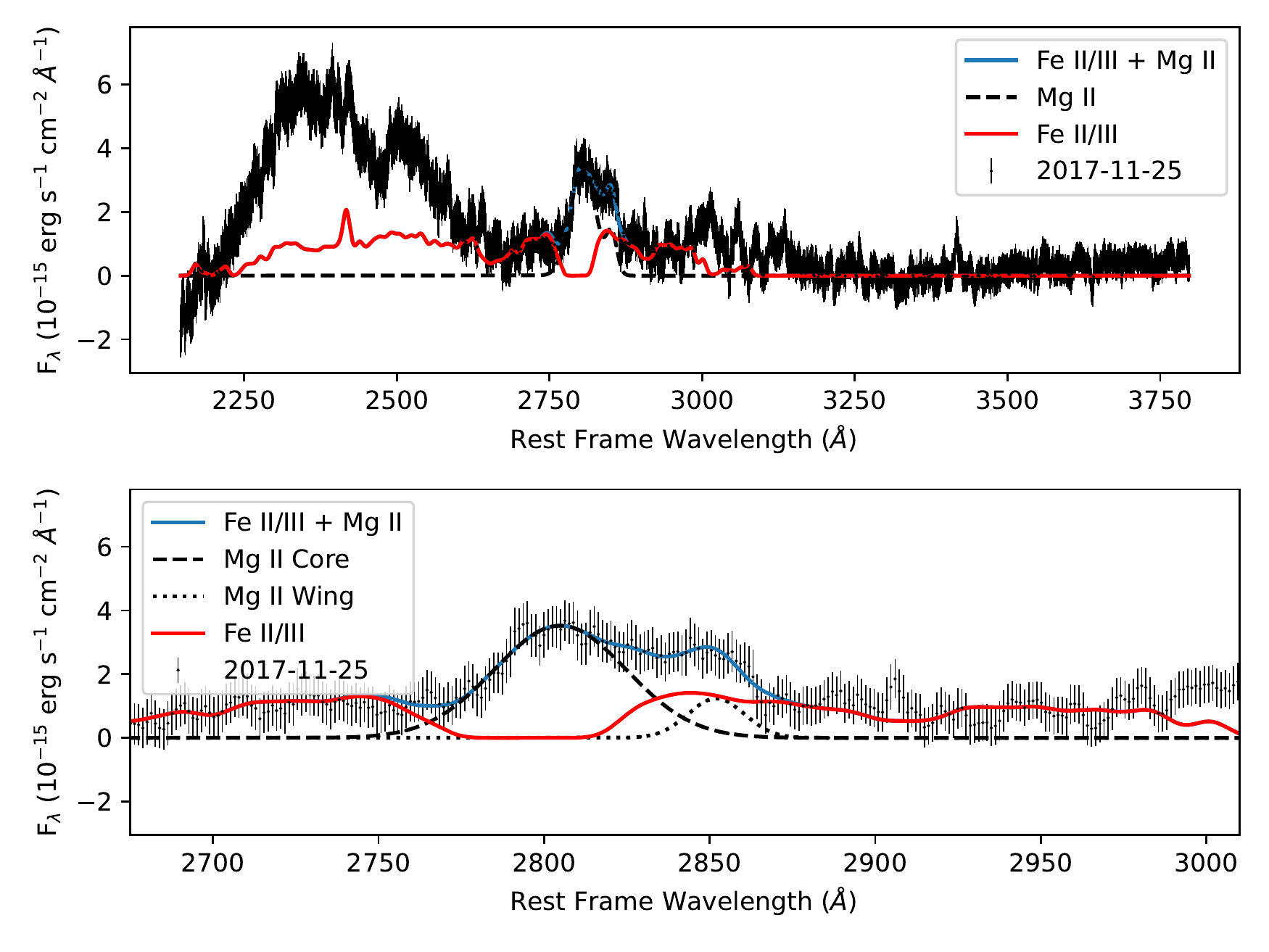}{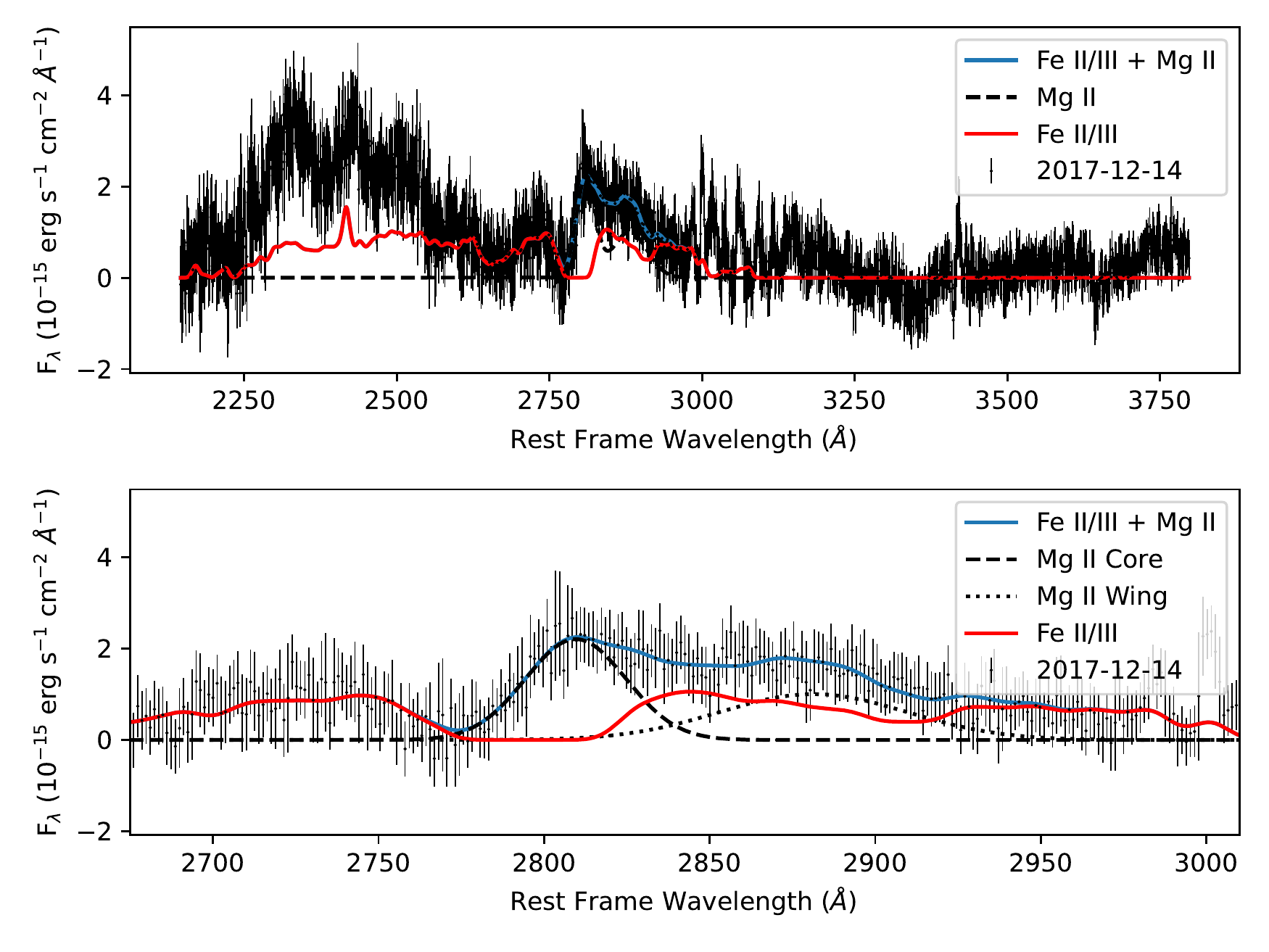}}\\
    {\plottwo{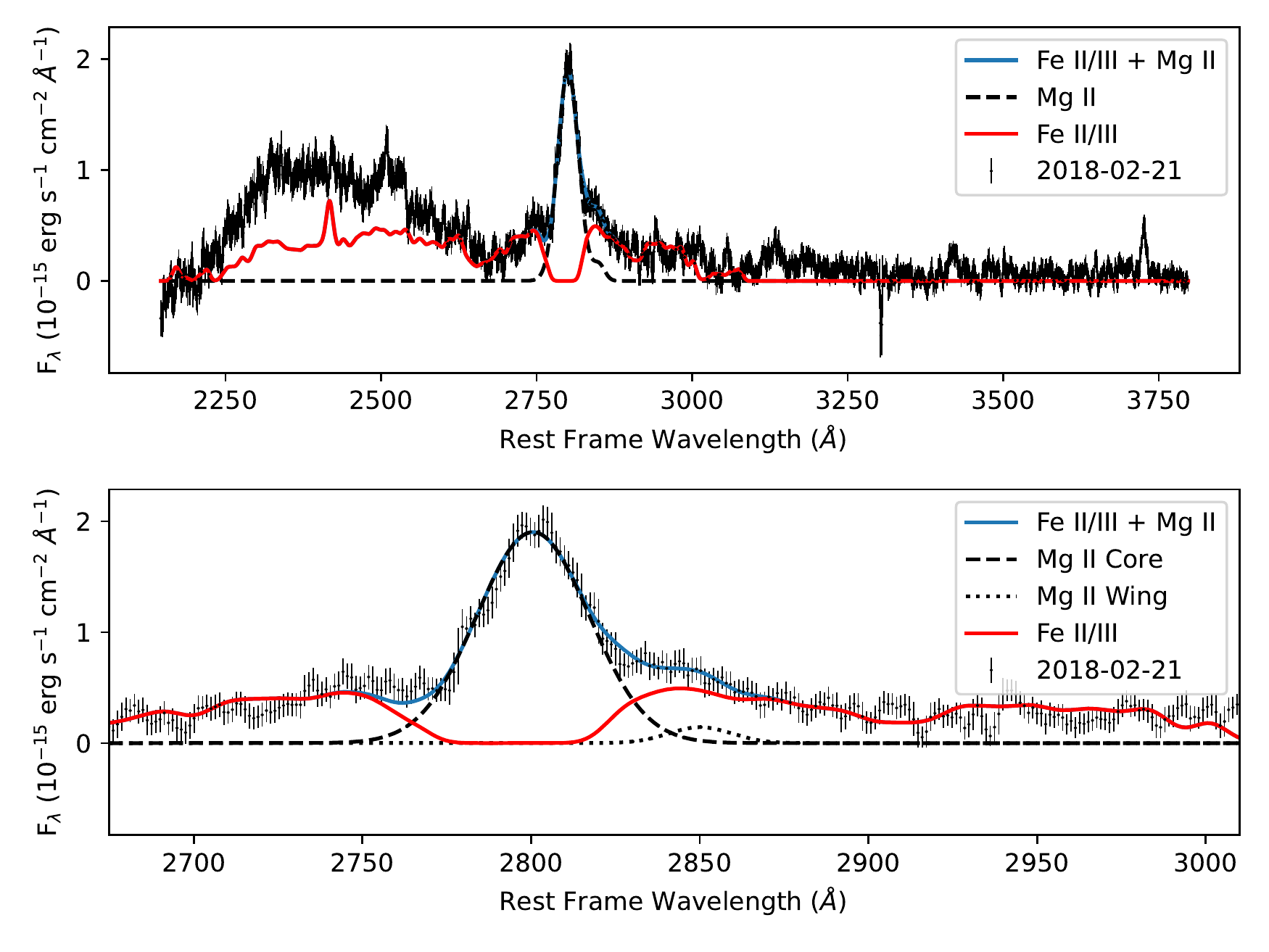}{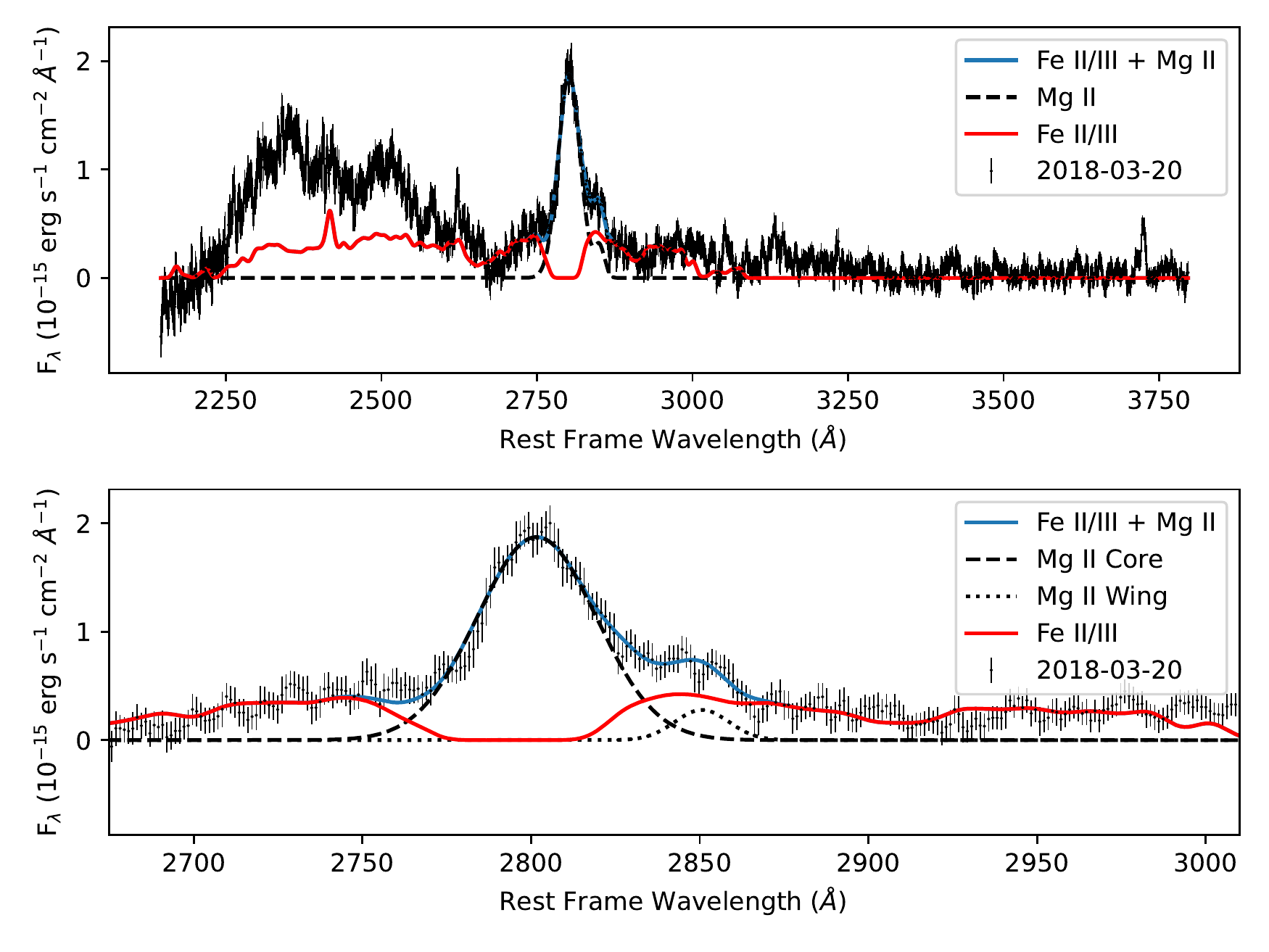}}
    \caption{Continuum-subtracted spectra plotted with the best-fit Gaussian \ion{Mg}{2} $\lambda2798$ emission line profiles and scaled Fe II/III template (see text).}
\end{figure*}
\begin{figure*}[!tp]
    \ContinuedFloat
    \centering
    {\plottwo{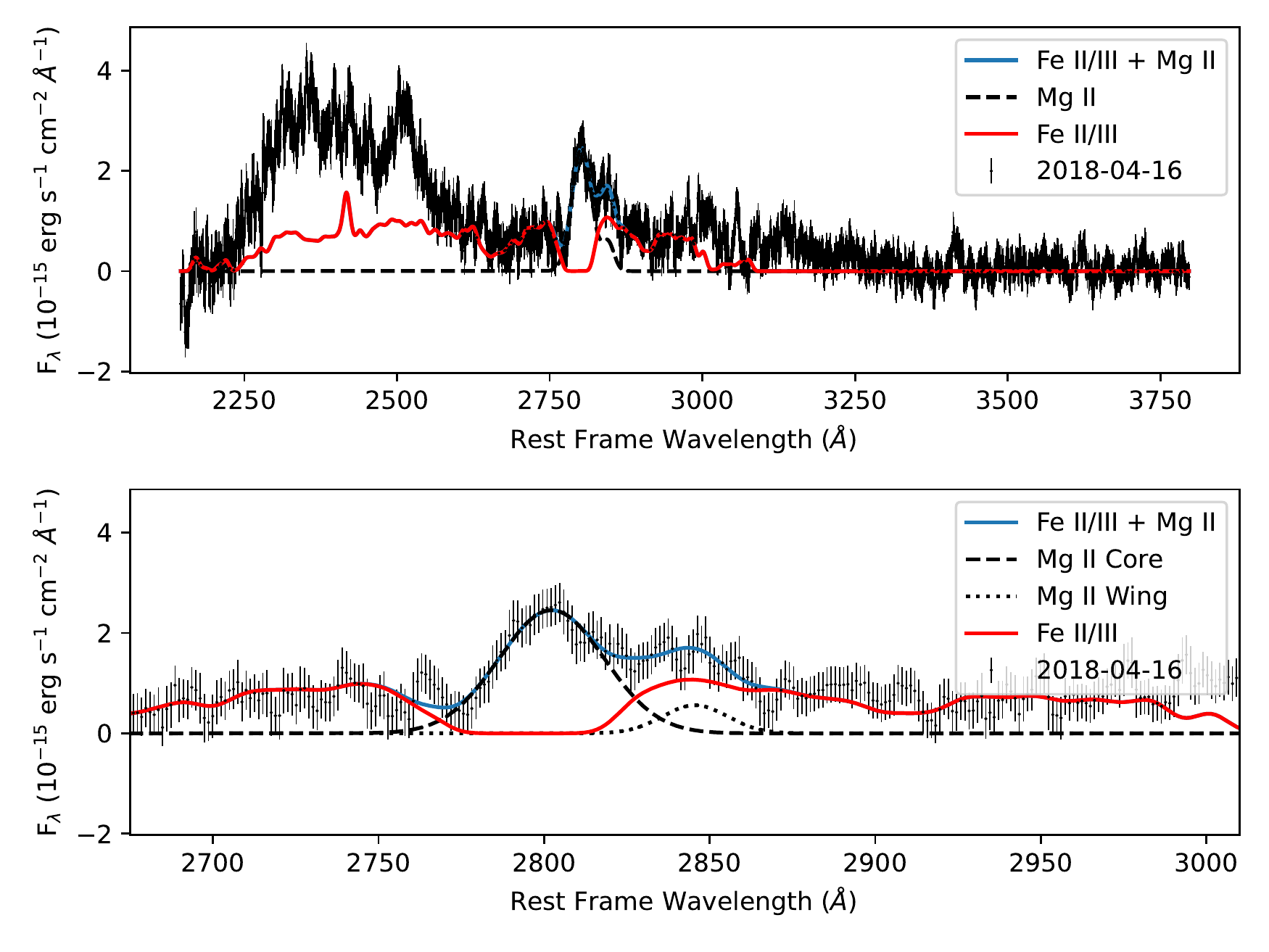}{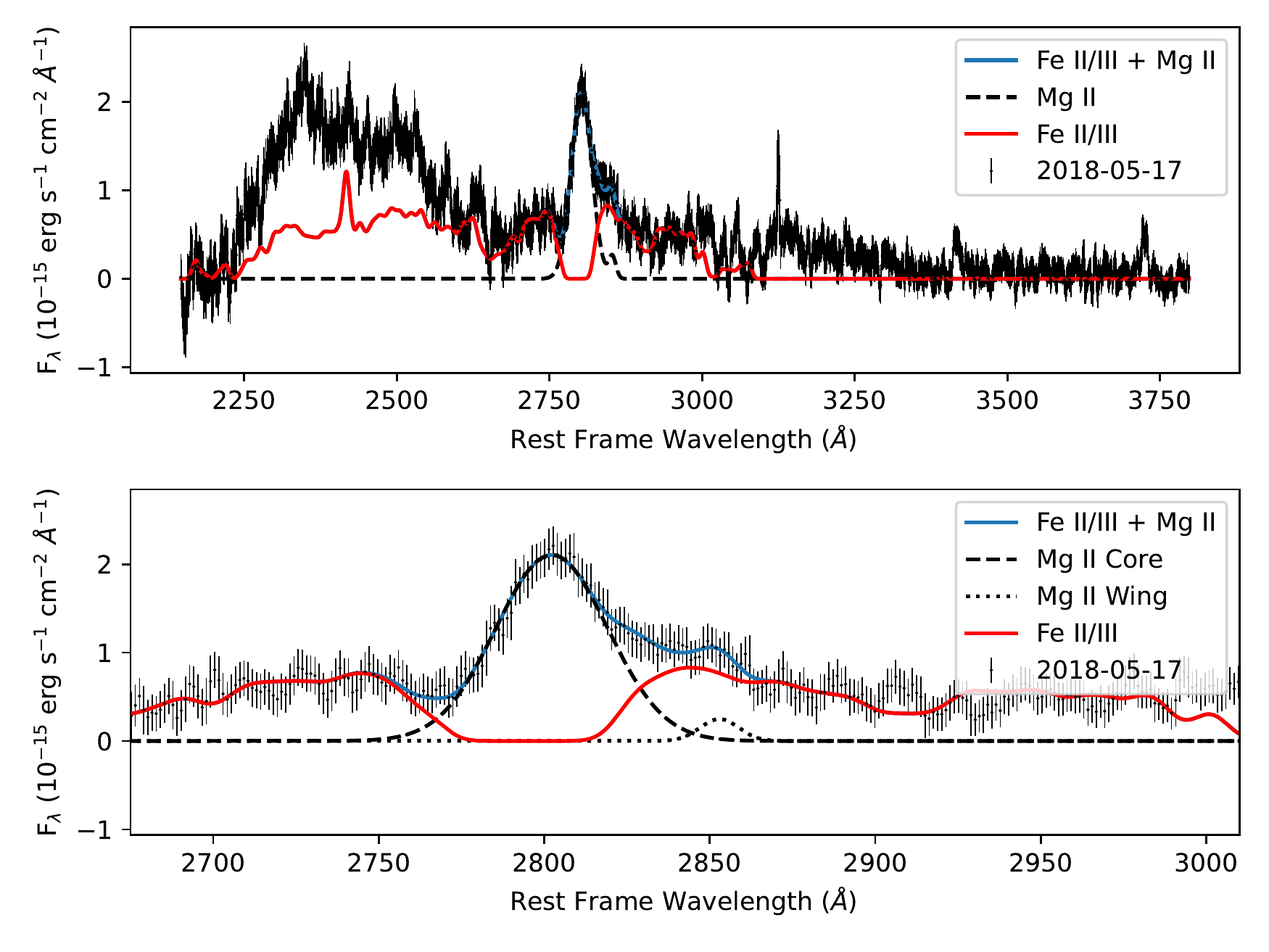}}\\
    {\plottwo{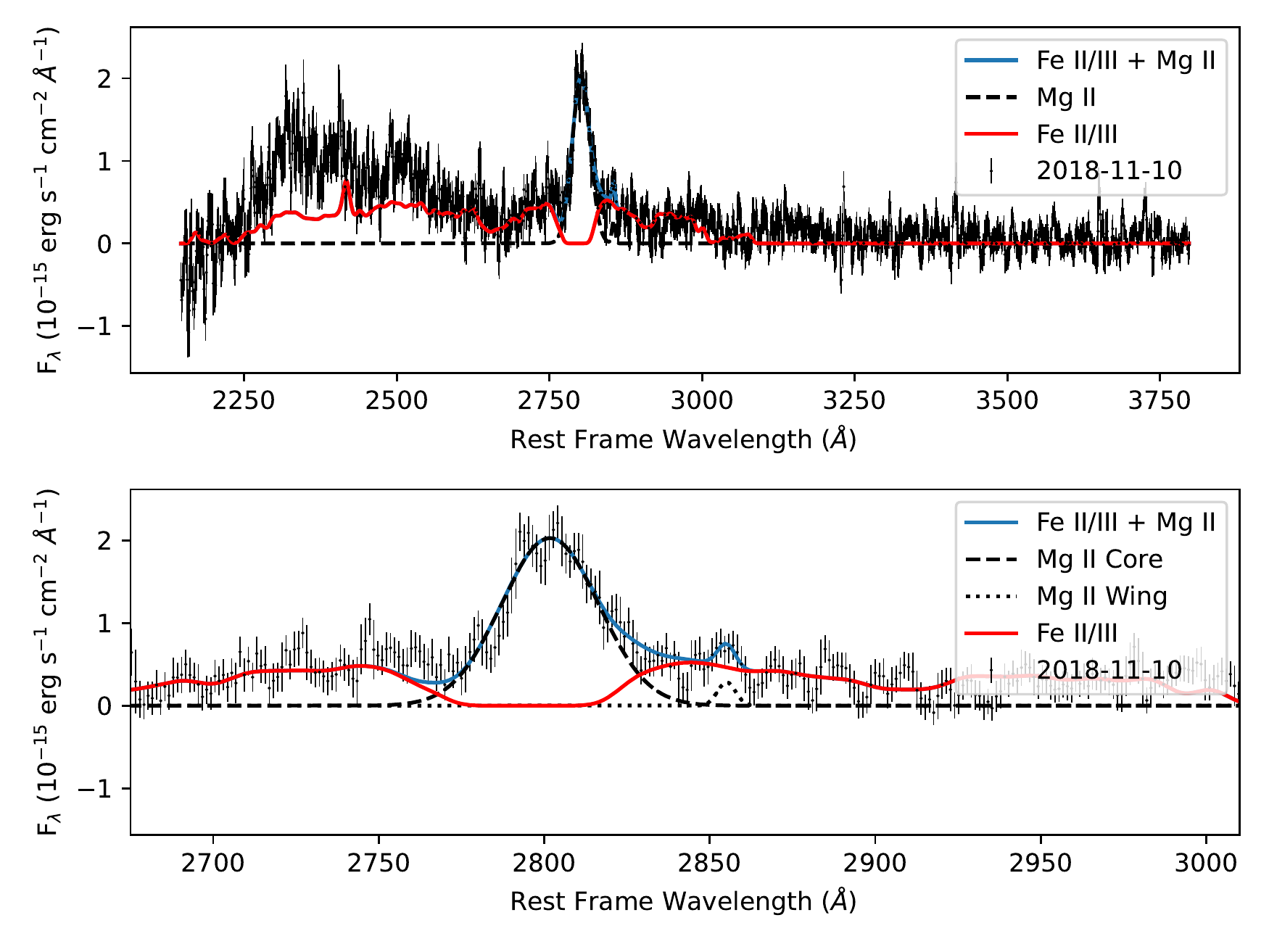}{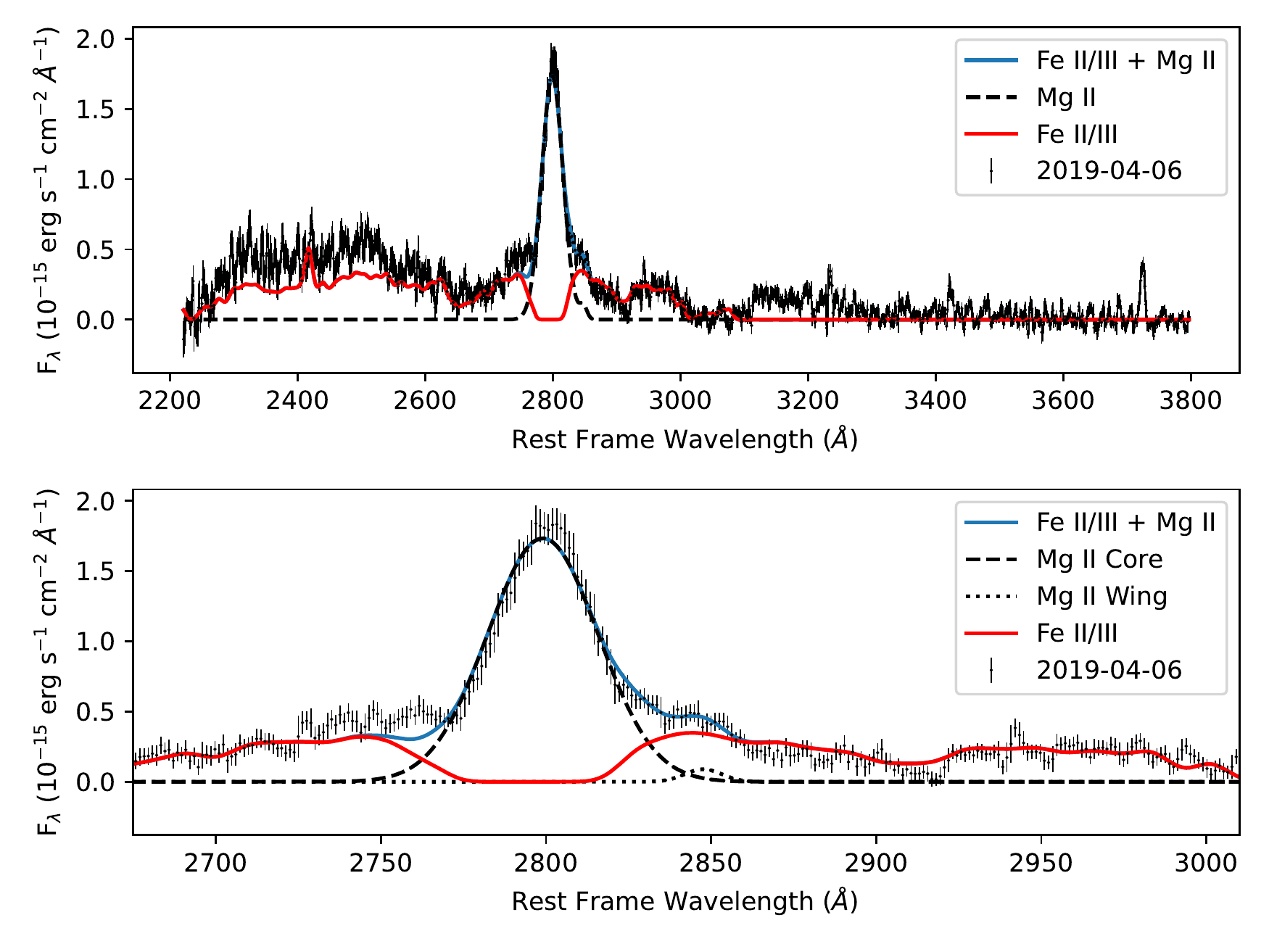}}\\
    \epsscale{0.5}
    {\plotone{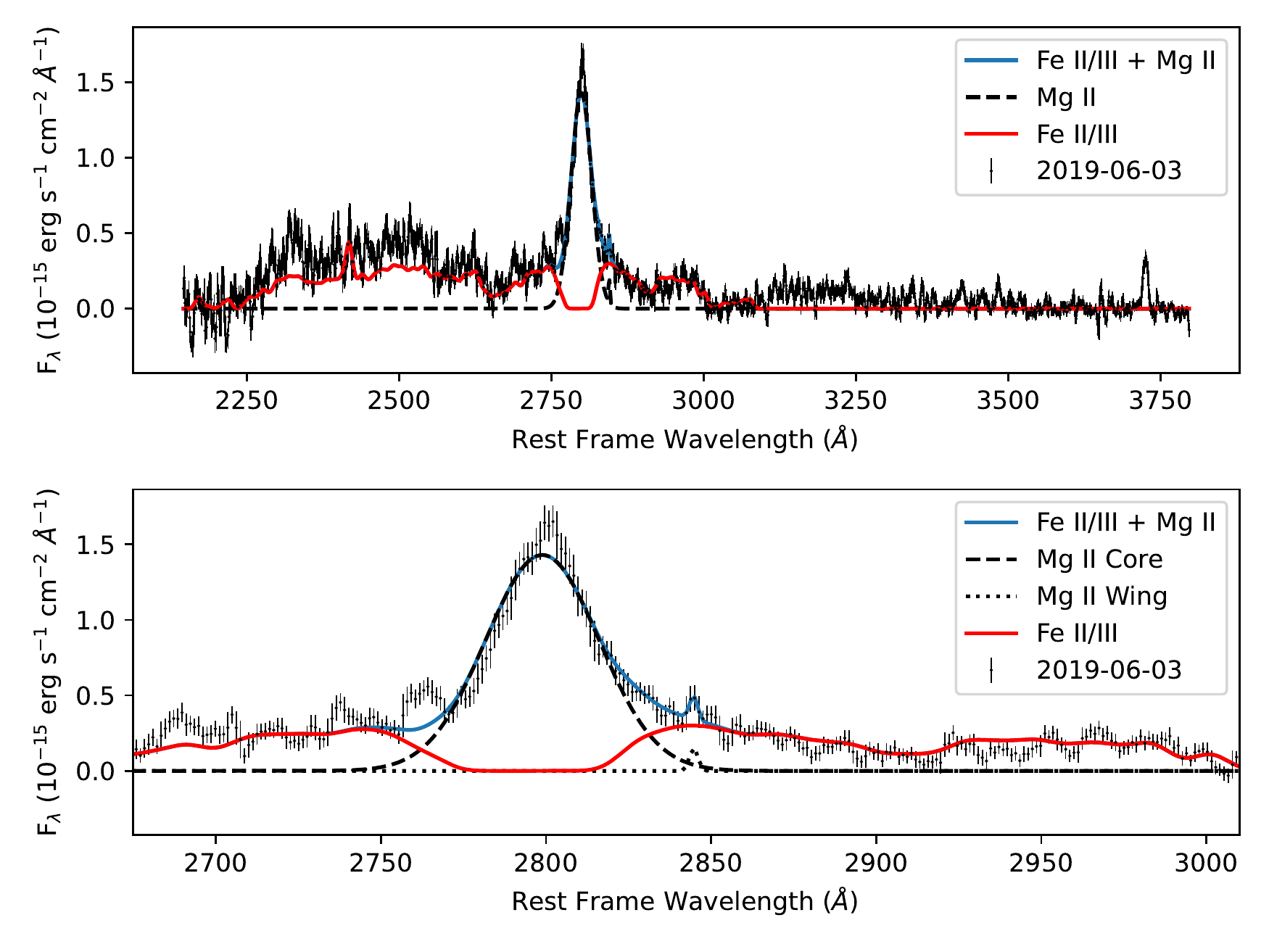}}\\
    \caption{Continued.}
    \label{fig06}
\end{figure*}

\subsection{Fe Emission}\label{sec:Fe}

\begin{figure}[!htb]
    \plottwo{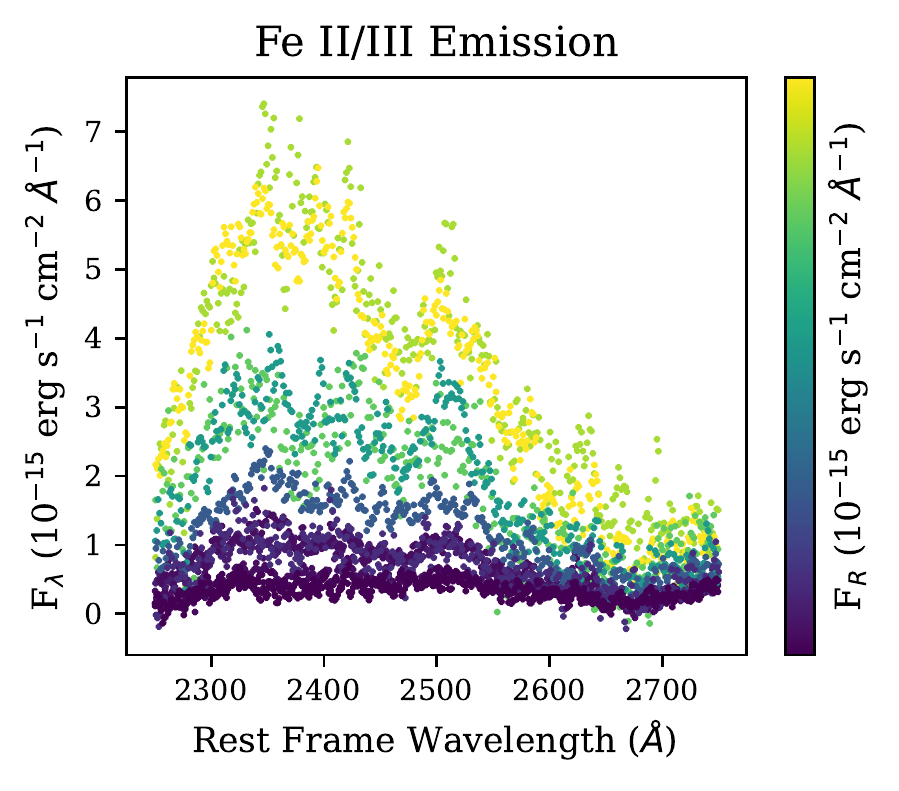}{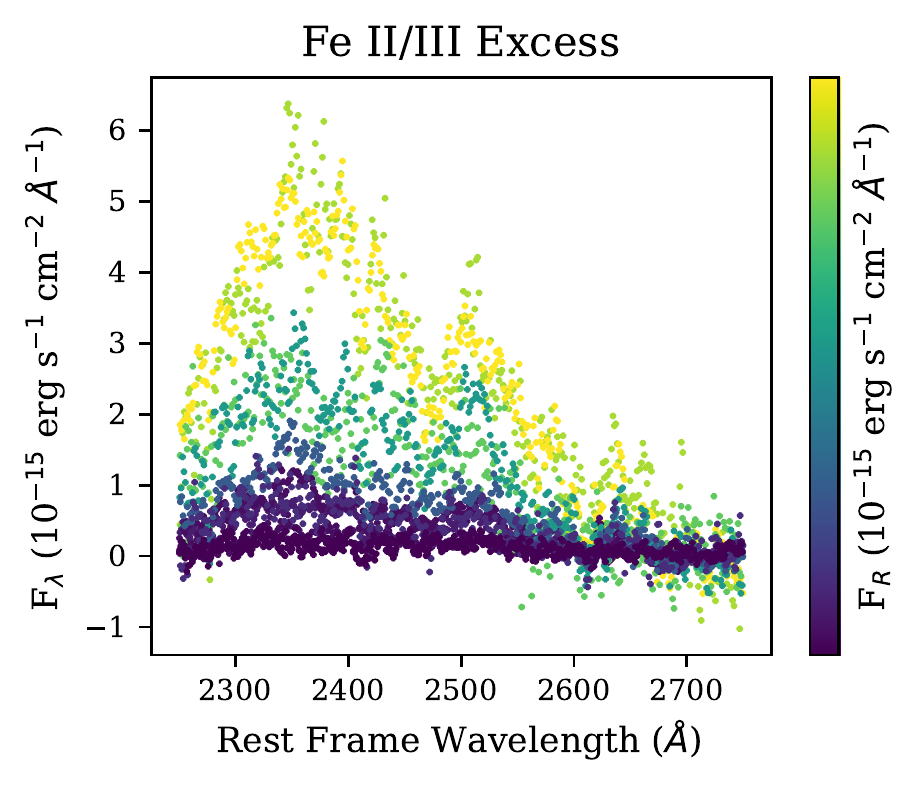}
    \caption{\emph{Left:} The continuum subtracted spectra in the 2250-2750 \AA\  region. (\emph{Right}) The Fe II/III excess, i.e., difference between the continuum subtracted spectra and the best-fit Fe templates. The color of each spectrum corresponds to the R band magnitude of the source converted to flux density (see color bar). (These spectra contain emission from non-Fe II/III lines as well; however, the Fe II/III lines dominate.)}
    \label{fig07}
\end{figure}

Figure~\ref{fig07} examines more closely the underestimation of the flux density of the broad emission feature blueward of the \ion{Mg}{2} $\lambda2798$ line. The left panel of the figure presents each spectrum with just the continuum profile subtracted, while the right panel shows the same spectrum with the corresponding broadened and scaled Fe template subtracted as well. As can be seen in the figure, the flux of the Fe emission complex increases with continuum flux, as does the residual after the fitted Fe template is subtracted. Table~\ref{tab6} reports the Fe emission flux as measured by integrating under the fitted template (column 2), the flux as measured by integrating the continuum from \FeLower\ to \FeUpper\ \AA\ (column 3), and the difference between these two values (column 4). The Fe template is based on the Seyfert 1 galaxy I Zwicky 1, while the strength of the Fe-complex lines is known to vary across different objects \citep[see][]{vestergaard2001}. The physical cause of this is poorly understood, and the general lack of knowledge about the main processes governing the strengths of the Fe lines leaves unexplained the discrepancy between the template and our spectra in the region blueward of the \ion{Mg}{2} $\lambda2798$ emission line.

\begin{deluxetable}{c C C C}[htb!]
    \tablecaption{\ion{Fe}{2} Emission Flux \label{tab6}}
    \tablehead{
        \colhead{Epoch} & 
        \dcolhead{F_{\textrm{Fe, template}}} & 
        \dcolhead{F_{\textrm{Fe, spec}}} \\ 
        \colhead{yyyy-mm-dd } & \dcolhead{\mathrm{perg\ s^{-1} cm^{-2}}} & \dcolhead{\mathrm{perg\ s^{-1} cm^{-2}}} \\
        \colhead{(1)} &
        \colhead{(2)} &
        \colhead{(3)}
        }
    \startdata
    2017-05-27 & 126\pm3 & 205\pm2 \\ 
    2017-11-20 & 579\pm23 & 1725\pm21 \\ 
    2017-11-25 & 481\pm15 & 1614\pm19 \\ 
    2017-12-14 & 360\pm18 & 906\pm21 \\ 
    2018-02-21 & 168\pm4 & 345\pm3 \\ 
    2018-03-20 & 144\pm5 & 361\pm4 \\ 
    2018-04-16 & 365\pm12 & 953\pm \\ 
    2018-05-17 & 282\pm6 & 602\pm6 \\ 
    2018-11-10 & 178\pm7 & 358\pm7 \\ 
    2019-04-06 & 119\pm3 & 183\pm2 \\ 
    2019-06-03 & 102\pm3 & 154\pm2 \\ 
    \enddata
    \tablecomments{Columns include: (1) UT date of observation; (2) Fe flux from integrating the template spectrum; (3) Fe flux from integrating the continuum and Mg II free spectrum over the wavelengths 2250-2750 \AA. (It should be noted that these reported fluxes do contain emission from non-Fe II/III lines, however, the Fe II/III lines dominate.)
    }
\end{deluxetable}
We have also performed the same fitting routine using an \ion{Fe}{2} template by \cite{popovic2018}, finding that the general behavior of the \ion{Mg}{2} $\lambda 2798$ line, as well as the increase of the \ion{Fe}{2} emission flux with the continuum, persist for this template as well. Additionally, the presence of a red wing component of \ion{Mg}{2} $\lambda 2798$ is confirmed using this template. This wing also tends to be brighter when the blazar is in an active state. Details regarding the analysis using this Fe template can be found in the appendix. 
In the discussion below we consider the results obtained with the template by \cite{vestergaard2001}.

\section{Discussion} \label{sec:discuss}

\begin{figure*}[!htb]
    \plotone{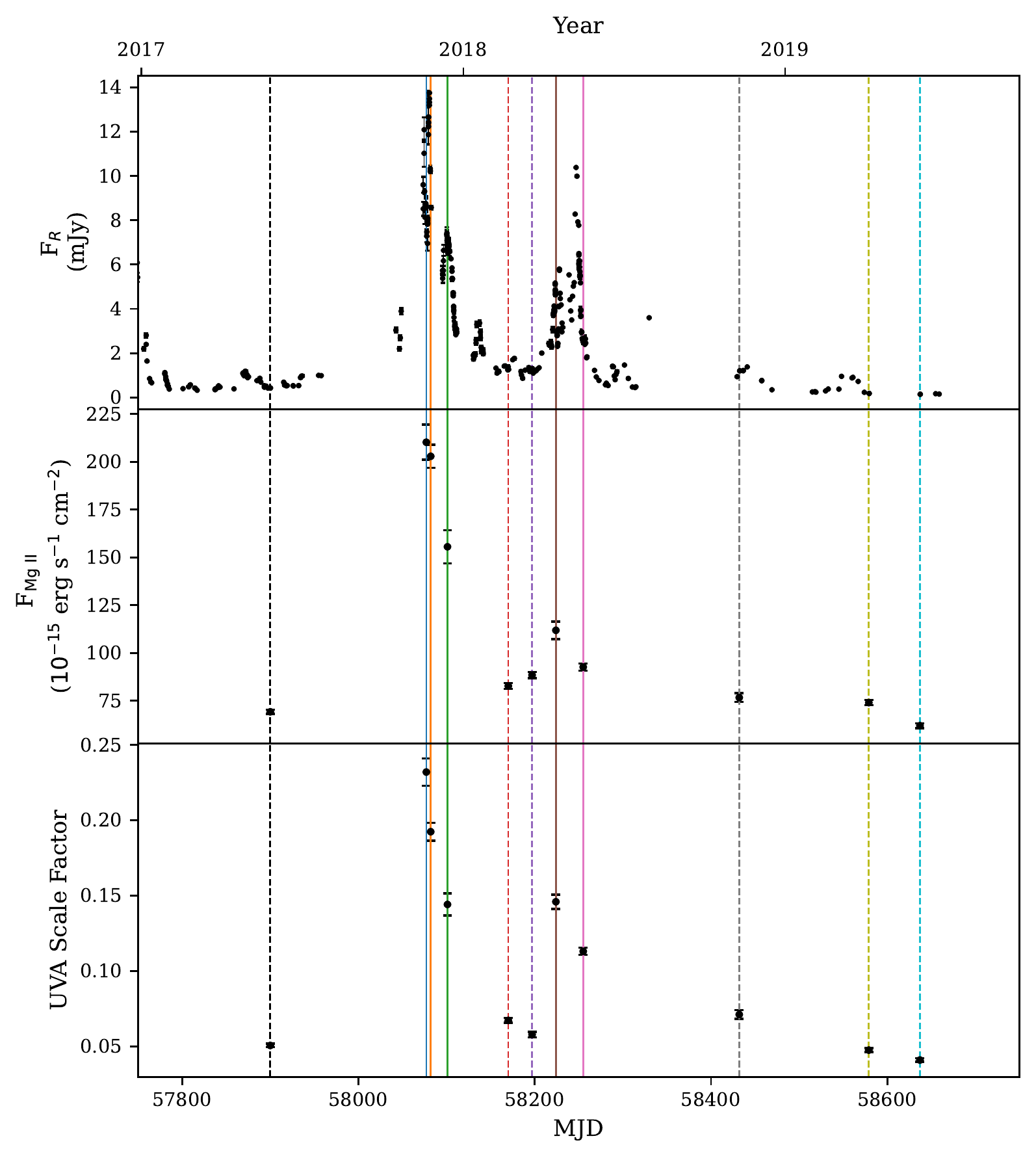}
    \caption{From top to bottom: optical continuum, \ion{Mg}{2} emission-line, and Fe II/III emission-line fluxes of \source\ as a function of time. Vertical lines mark times of the spectroscopic observations.}
    \label{fig08}
\end{figure*}

\begin{figure*}[!htb]
    \plottwo{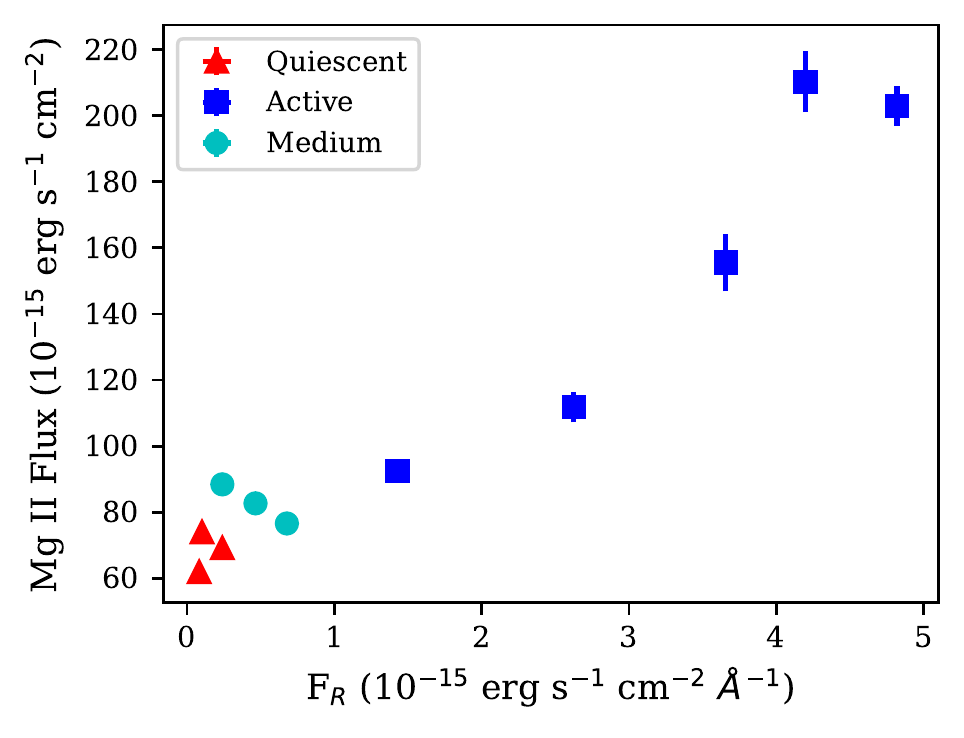}{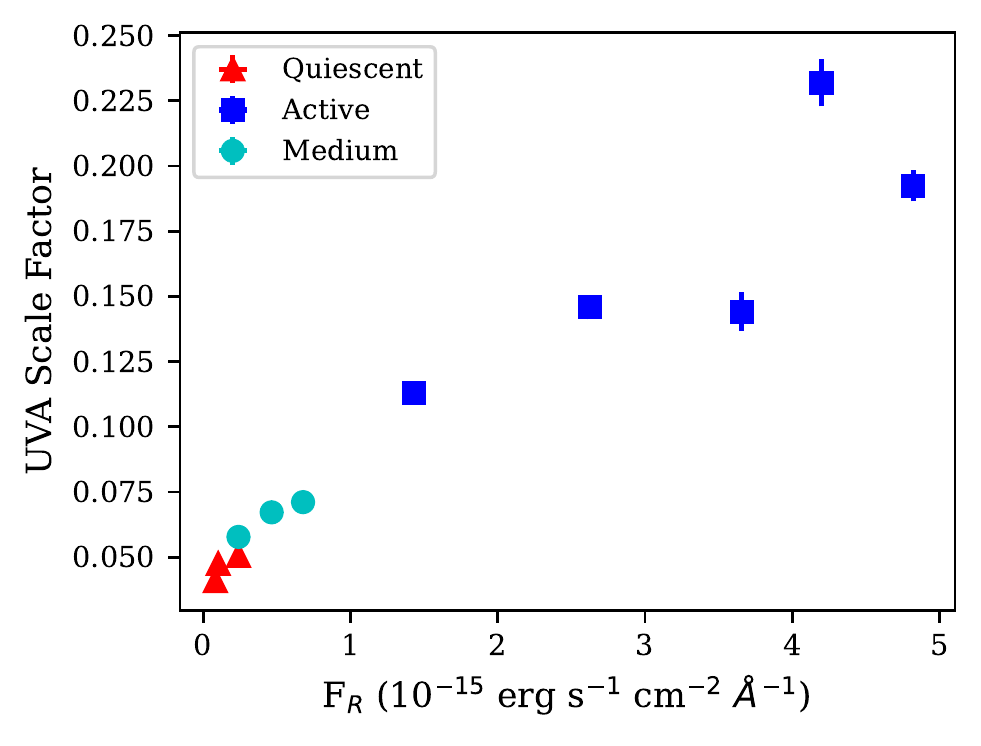}
    \caption{Emission line fluxes of \ion{Mg}{2} (\emph{left}) and \ion{Fe}{2} (\emph{right}) versus the R band flux density.}
    \label{fig09}
\end{figure*}

The variations of the continuum, \ion{Mg}{2} line, and Fe II/III complex fluxes are displayed in Figures~\ref{fig08} and \ref{fig09}. The maximum total luminosity of the \ion{Mg}{2} plus Fe II/III lines is $\sim7\times10^{45}$ erg~s$^{-1}$, with the latter $\sim9$ times higher than the former.
As can be seen in both figures, the \ion{Mg}{2} and Fe II/III fluxes rise and fall with the continuum flux.

Although there are too few spectroscopic measurements to perform a robust correlation analysis, the near-simultaneity of the maximum observed line fluxes and maximum optical continuum flux strongly suggests a connection with a time delay $\lesssim2$ weeks.
This implies that UV photons from the jet ionize the emission line producing clouds, which respond to variations in the jet flux with a short observed time delay, $\lesssim2$ weeks (see Figure  \ref{fig08}). This can only be explained if the clouds producing the emission lines are located near the jet.  We consider a cloud at a distance $r$ from the source of the continuum flare in the jet and at a distance $x$ from the jet axis. Changes in emission lines relative to continuum photons traveling directly along the line of sight from the flare site will occur with a light-travel time delay, $\Delta t$, governed by the expression
\begin{equation}
x \approx [c\Delta t^{'}(c\Delta t^{'}+2r)]^{1/2},
\end{equation}
where $\Delta t^{'}=\Delta t_{\text{obs}}(1+z)^{-1}$ is measured in the rest frame of the host galaxy. This equation is valid for small viewing angles $\theta$ of the jet, such that $x\gg r\tan\theta$, as is the case for \source\ ($\theta\leq2.5\degr$). For an observed time delay $\Delta t_{\text{obs}}\sim2$ weeks, $x \approx 0.1r^{1/2}$ pc. This places the clouds in the environment immediately surrounding the parsec-scale jet.

The shift of the center of the \ion{Mg}{2} line toward longer wavelengths during high flux states
implies that the contribution to the increase in line flux was greater for clouds receding more rapidly from us. This can be explained if the clouds are falling toward the SMBH. This is similar to, but not as large as, the shift seen in variations of the \ion{Mg}{2} line in the blazar 3C~279. The latter included the appearance of a strong red wing to the line, centered $\sim3500$ km~s$^{-1}$ from the center of the pre-flare line \citep{punsly2013,larionov2020}.

If the velocity shift were due to the free-fall motion of the clouds in the gravitational potential of the SMBH, the distance from the SMBH of the clouds producing the red wing would be 
\begin{equation}\label{eq:Rupper}
    R_{\rm{ff}} = 2R_{\rm{g}}(c/v)^2.
\end{equation}
Here, $R_{\rm{g}}= GM/c^2$ is the gravitational radius,
where $G$ is the gravitational constant and $M$ is the mass of the SMBH.
The observed shift in the red wing is $\sim5000$ km s$^-1$ for all spectra. 
The free-fall distance of the red-wing clouds is therefore $\sim7200 R_{\rm{g}}$. This is within the sphere of influence of the black hole \citep{zhou2019}. The gravitational radius of the SMBH is $4.3 \times 10^{-5}$ pc for our adopted mass of $9\times10^8 M_\odot$. The distance of the red wing clouds would therefore need to lie within the inner 0.3 pc. However, the flare occurred simultaneously with a detection of VHE photons from the quasar, which cannot escape from a region so close to the SMBH, as discussed in \S\ref{sec:intro}. We therefore conclude that the excess redshift of these clouds is not caused by gravitational infall.

\begin{figure}[!htb]
    \plotone{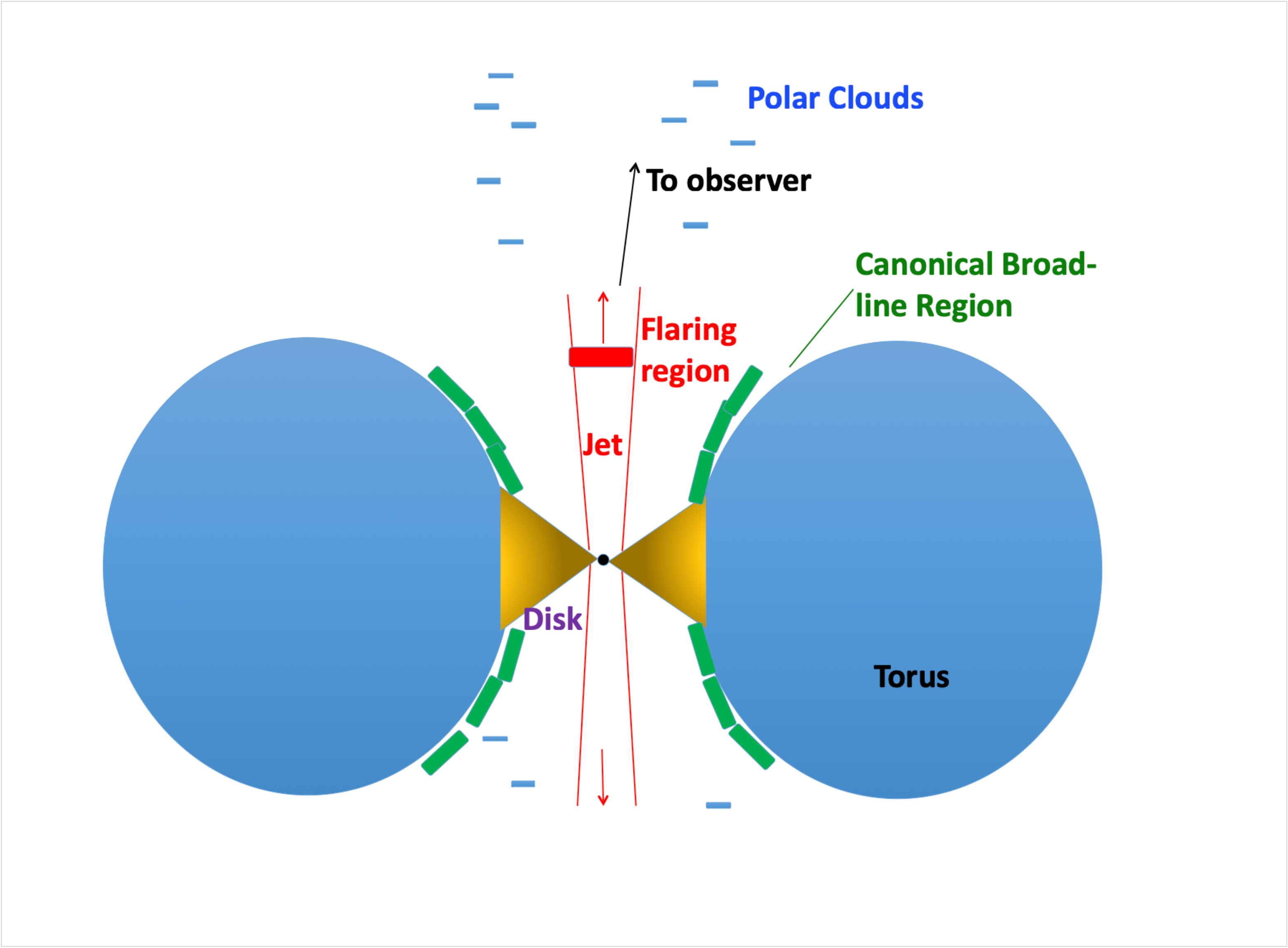}
    \caption{Sketch of the possible locations of the main components of the quasar 1156+295, which may represent that of blazars in general.}
    \label{fig10}
\end{figure}

The origin and kinematics of the infalling polar clouds are thus challenging to explain.
One possibility is that the clouds are entrained in a backflow from the interaction between the terminus of the jet and the kiloparsec-scale interstellar medium \citep[which can flow toward the accretion disk at a speed as high as $\sim20,000$ km~s$^{-1}$; e.g.,][]{Cielo2014,Perucho2019, norman1984}. Another proposal is that the clouds (as well as non-polar emission-line gas) are condensations within an accretion-disk wind, dense regions that fail to reach the escape velocity, causing them to fall back toward the SMBH \citep{Elvis2017}. Such polar clouds should be optically thin, otherwise the line radiation along the line of sight --- which lies within 2.5\degr\ of the jet axis --- would be shielded. However, the free-fall velocity of the condensations  at $r\sim6$ pc would be much less than the observed values of $\sim5000$ km~s$^{-1}$ (see Table \ref{tab5}).

The symmetry of the observed emission lines in \source\ is difficult to reproduce with a model based on pure inflow. If all of the red-wing clouds were moving toward the SMBH, the \ion{Mg}{2} line profile would be skewed toward the long or short wavelength side, depending on whether the radial velocity increases toward larger or smaller distances $r$ from the SMBH. The velocities of the clouds could have a turbulent component that accounts for most of the line width. Such turbulence would be natural for clouds embedded in a backflow, caused by Kelvin-Helmholtz instabilities at the interface with the jet \citep{Cielo2014,Perucho2019}. The turbulence should be stronger closer to the jet, which could account for the increased breadth of the \ion{Mg}{2} line during the outburst in \source, to which clouds closest to the jet respond first.

The existence of broad-line emitting clouds near the jet provides a possible source of seed photons for $\gamma$-ray production. (The observed ratio of maximum $\gamma$-ray to optical luminosity of $\sim10$ implies that EC rather than SSC scattering dominates the $\gamma$-ray production; see \S\ref{sec:intro}.) As discussed in \S\ref{sec:intro}, VHE emission produced in the sub-parsec region around the SMBH is suppressed due to pair production \citep{aleksic2011}.
The quasar was detected at energies $E>100$ GeV on 2017 December 12 \citep{mirzoyan2017}, $\sim2$ weeks
after we measured the maximum flux of the \ion{Mg}{2} line and \ion{Fe}{2} complex (see Tables~\ref{tab5} and \ref{tab6}). Note also that during these two weeks a significant increase in the degree of optical polarization, up to $\sim22\%$, was observed (see Figure~\ref{fig01}), confirming that the optical flare was nonthermal, and hence represents beamed emission from the jet. Therefore, if emission-line clouds near the axis of the jet are providing the seed photons, they must also be located beyond the inner parsec-sized region. If the clouds producing the variable emission lines are located at $r\sim6$ pc, as in the wind condensation model discussed above, the energy density of the line photons at this distance is $\sim5\times10^{-3}$ erg~cm$^{-3}$ in the rest-frame of the jet. This is sufficient to provide the seed photons needed to explain a typical $\sim1$ GeV $\gamma$-ray flare \citep[e.g.,][]{joshi2014}.

\section{Conclusions} \label{sec:conclude}

Based on our spectroscopic and continuum flux monitoring, we conclude that polar clouds, situated alongside the jet and located parsecs from the central SMBH, are ionized by UV photons from flares of synchrotron radiation originating farther upstream in the jet. This increases the flux of the \ion{Mg}{2} $\lambda2798$ \AA\ line and \ion{Fe}{2} emission complex, with a time delay $\lesssim2$ weeks. We sketch the inferred location of this extended broad-line region in Figure~\ref{fig10}.

The variations of the \ion{Mg}{2} line are similar to, but less extreme than, those seen in 3C~279 \citep{punsly2013,larionov2020}, with the line broadening and the central wavelength shifting to longer values when the continuum flux is higher. In order to sample the range of behavior of the emission lines in blazars, we plan to continue to follow changes in the emission lines of \source, as well as those of other $\gamma$-ray flaring blazars. 

\acknowledgments
{We thank the referee for contructive comments that improved the paper. We are grateful to M.\ Vestergaard for providing the Fe emission template. This research was supported in part by NASA Fermi guest investigator program grants 80NSSC19K1504 and 80NSSC20K1565. We thank A.\ Tchekhovskoy for discussion of possible origins of the variable line-emitting clouds. These results made use of the Lowell Discovery Telescope (LDT) at Lowell Observatory. Lowell Observatory is a private, non-profit institution dedicated to astrophysical research and public appreciation of astronomy, and operates the LDT in partnership with Boston University, the University of Maryland, the University of Toledo, Northern Arizona University and Yale University. This study was based in part on observations conducted using the 1.8 m Perkins Telescope Observatory (PTO) in Arizona, which is owned and operated by Boston University. I.\ A. acknowledges financial support from the Spanish ``Ministerio de Ciencia e Innovaci\'on'' (MCINN) through the ``Center of Excellence Severo Ochoa'' award for the Instituto de Astrof\'isica de Andaluc\'ia-CSIC (SEV-2017-0709). Acquisition and reduction of the MAPCAT data were supported in part by MICINN through grants AYA2016-80889-P and PID2019-107847RB-C44. The MAPCAT observations were carried out at the German-Spanish Calar Alto Observatory, which is jointly operated by Junta de Andaluc\'ia and Consejo Superior de Investigaciones Cient\'ificas. Data from the Steward Observatory spectropolarimetric monitoring project were used; this program was supported by Fermi Guest Investigator grants NNX08AW56G, NNX09AU10G, NNX12AO93G, and NNX15AU81G. C.C. acknowledges support from the European Research Council (ERC) under the European Union Horizon 2020 research and innovation program under the grant agreement No 771282.}

\facilities{LDT (DeVeny, LMI), Perkins, Fermi}

\software{
    Astropy, \citep{astropy:2013, astropy:2018},
    HyperSpy, \citep{hyperspy}
    LMFIT \citep{newville2014}
    Matplotlib \citep{hunter2007},
    NumPy \citep{harris2020},
    Specutils \citep{specutilsteam2021},
    uncertainties \citep{lebigot},
    SciPy \citep{scipy2020},
    SQLAlchemy \cite{sqlalchemy},
    pandas \citep{reback2020},
    PyCCF \citep{peterson1998, sun2018}
    }

\appendix

The \ion{Fe}{2} template created by \cite{popovic2018} is a composite of the  \ion{Fe}{2} multiplets 60, 61, 62, 63, and 78, as well as  additional \ion{Fe}{2} emission lines found in the quasar I Zw 1. The template covers the rest frame wavelength range 2650-3050 \AA. Each component is scaled arbitrarily relative to each other. We used the 3000 km/s FWHM broadened template and employed the same fitting procedures as we did when using the template from \cite{vestergaard2001}. The fitted parameters are reported in Table~\ref{tab7}, and are plotted in Figure~\ref{fig11}. The fitted \ion{Mg}{2} flux and \ion{Fe}{2} flux versus the R band flux density are plotted in Figure~\ref{fig12}. The \ion{Mg}{2} emission and the \ion{Fe}{2} emission increase with the continuum as in the case of usage of the template by \cite{vestergaard2001} (see Figure~\ref{fig09}).

\begin{splitdeluxetable*}{@{\extracolsep{4pt}} c C C C C C C B C C C C C C C@{}}
    \tablecolumns{14}
    \tablecaption{Results of \ion{Mg}{2} $\lambda2798$ Emission Line Fitting When Using the Fe II Template from \cite{popovic2018} \label{tab7}}
    \tablehead{
        \colhead{Epoch} &
        \multicolumn{2}{c}{Shift} &
        \multicolumn{2}{c}{FWHM} &
        \multicolumn{2}{c}{Flux} &
        \multicolumn{6}{c}{Fe Template Scale} &
        \colhead{$\chi^2_{\rm dof}$}
        \\ 
        \colhead{yyyy-mm-dd} & 
        \multicolumn{2}{c}{$\mathrm{km\,s^{-1}}$} & 
        \multicolumn{2}{c}{$\mathrm{km\,s^{-1}}$} & 
        \multicolumn{2}{c}{$10^{-15}\mathrm{erg\,s^{-1}\,cm^{-2}}$} & 
        \colhead{} &
        \colhead{} &
        \colhead{} &
        \colhead{} &
        \colhead{} &
        \colhead{} &
        \colhead{}
        \\
        \cline{2-3}
        \cline{4-5}
        \cline{6-7}
        \cline{8-13}
        &
        \colhead{Core} &
        \colhead{Wing} &
        \colhead{Core} & 
        \colhead{Wing} &  
        \colhead{Core} &
        \colhead{Wing} &
        \colhead{Mul. 60} &
        \colhead{Mul. 61} &
        \colhead{Mul. 62} &
        \colhead{Mul. 63} &
        \colhead{Mul. 78} &
        \colhead{IZw1lines} &
        \colhead{}
        \\
        \colhead{(1)} &
        \colhead{(2)} &
        \colhead{(3)} &
        \colhead{(4)} &
        \colhead{(5)} &
        \colhead{(6)} &
        \colhead{(7)} &
        \colhead{(8)} &
        \colhead{(9)} &
        \colhead{(10)} &
        \colhead{(11)} &
        \colhead{(12)} &
        \colhead{(13)} &
        \colhead{(14)}
        \\
        }
    \startdata
    2017-05-27 &   178 \pm  35 &  ... &  3680 \pm  77 & ... &   64.1 \pm  1.2 & ... &   0.016 \pm  0.001 &   0.015 \pm  0.001 &   0.020 \pm  0.003 &   0.000 \pm  0.007 &   0.016 \pm  0.002 &   0.017 \pm  0.002 &  2.399 \\
     2017-11-20 &   880 \pm 142 &  6401 \pm 149 &  4406 \pm 394 &   811 \pm 365 &  164.6 \pm 12.0 &   10.7 \pm  4.5 &   0.059 \pm  0.013 &   0.075 \pm  0.011 &   0.064 \pm  0.026 &   0.004 \pm  0.060 &   0.149 \pm  0.018 &   0.125 \pm  0.013 &  1.538 \\
     2017-11-25 &   573 \pm 119 &  6390 \pm 147 &  4827 \pm 367 &  1115 \pm 356 &  167.7 \pm 10.2 &   13.3 \pm  4.2 &   0.049 \pm  0.011 &   0.071 \pm  0.009 &   0.022 \pm  0.022 &   0.088 \pm  0.053 &   0.128 \pm  0.015 &   0.075 \pm  0.015 &  1.083 \\
     2017-12-14 &  1182 \pm 134 &  6665 \pm 179 &  3591 \pm 335 &  1082 \pm 420 &   81.1 \pm  6.3 &    9.0 \pm  3.4 &   0.018 \pm  0.012 &   0.012 \pm  0.020 &   0.001 \pm  0.016 &   0.129 \pm  0.040 &   0.082 \pm  0.012 &   0.053 \pm  0.011 &  0.669 \\
     2018-02-21 &   308 \pm  41 &  6747 \pm 263 &  3972 \pm 108 &  1536 \pm 631 &   76.4 \pm  1.7 &    2.6 \pm  1.1 &   0.020 \pm  0.002 &   0.023 \pm  0.002 &   0.026 \pm  0.001 &   0.000 \pm  0.000 &   0.024 \pm  0.003 &   0.019 \pm  0.003 &  1.580 \\
     2018-03-20 &   390 \pm  42 &  6321 \pm 109 &  4146 \pm 114 &   887 \pm 257 &   78.7 \pm  1.8 &    2.6 \pm  0.7 &   0.015 \pm  0.002 &   0.019 \pm  0.002 &   0.021 \pm  0.005 &   0.007 \pm  0.011 &   0.032 \pm  0.003 &   0.017 \pm  0.003 &  1.286 \\
     2018-04-16 &   427 \pm  87 &  6493 \pm 157 &  3909 \pm 229 &   622 \pm 372 &   95.2 \pm  4.6 &    3.2 \pm  1.7 &   0.043 \pm  0.007 &   0.059 \pm  0.005 &   0.041 \pm  0.013 &   0.003 \pm  0.032 &   0.093 \pm  0.010 &   0.066 \pm  0.005 &  1.333 \\
     2018-05-17 &   443 \pm  67 &  6444 \pm 250 &  4122 \pm 183 &  1254 \pm 577 &   86.7 \pm  3.1 &    3.8 \pm  1.8 &   0.034 \pm  0.004 &   0.038 \pm  0.003 &   0.025 \pm  0.008 &   0.035 \pm  0.018 &   0.053 \pm  0.006 &   0.037 \pm  0.004 &  1.504 \\
     2018-11-10 &   449 \pm  56 &  6171 \pm 118 &  3598 \pm 139 &   725 \pm 294 &   73.2 \pm  2.4 &    2.7 \pm  1.0 &   0.014 \pm  0.003 &   0.022 \pm  0.003 &   0.025 \pm  0.007 &   0.016 \pm  0.015 &   0.039 \pm  0.005 &   0.019 \pm  0.003 &  1.230 \\
     2019-04-06 &   160 \pm  38 &  ...          &  3858 \pm  87 & ...           &   69.1 \pm  1.4 &      ...        &   0.013 \pm  0.001 &   0.011 \pm  0.001 &   0.022 \pm  0.003 &   0.002 \pm  0.007 &   0.022 \pm  0.002 &   0.016 \pm  0.002 &  1.926 \\
     2019-06-03 &   138 \pm  49 &  ...          &  3954 \pm 120 & ...           &   58.6 \pm  1.5 &     ...         &   0.010 \pm  0.001 &   0.012 \pm  0.001 &   0.015 \pm  0.003 &   0.007 \pm  0.007 &   0.019 \pm  0.002 &   0.015 \pm  0.002 &  2.452 \\
    \enddata
    \tablecomments{Columns correspond to: (1) UT date of observation; (2-3) shift of the center of the \ion{Mg}{2} measured line from 2798 \AA\ rest wavelength for the core and wing component respectively; (4-5) FWHM of line profile for the core and wing component of \ion{Mg}{2} respectively; (6-7) flux of the core and wing components of Mg II line respectively; (8-13) scale factors used for the Fe II template (14) reduced $\chi^2$ of the fit. 
    }
\end{splitdeluxetable*}

\begin{figure*}[!bp]
    \centering
    {\plottwo{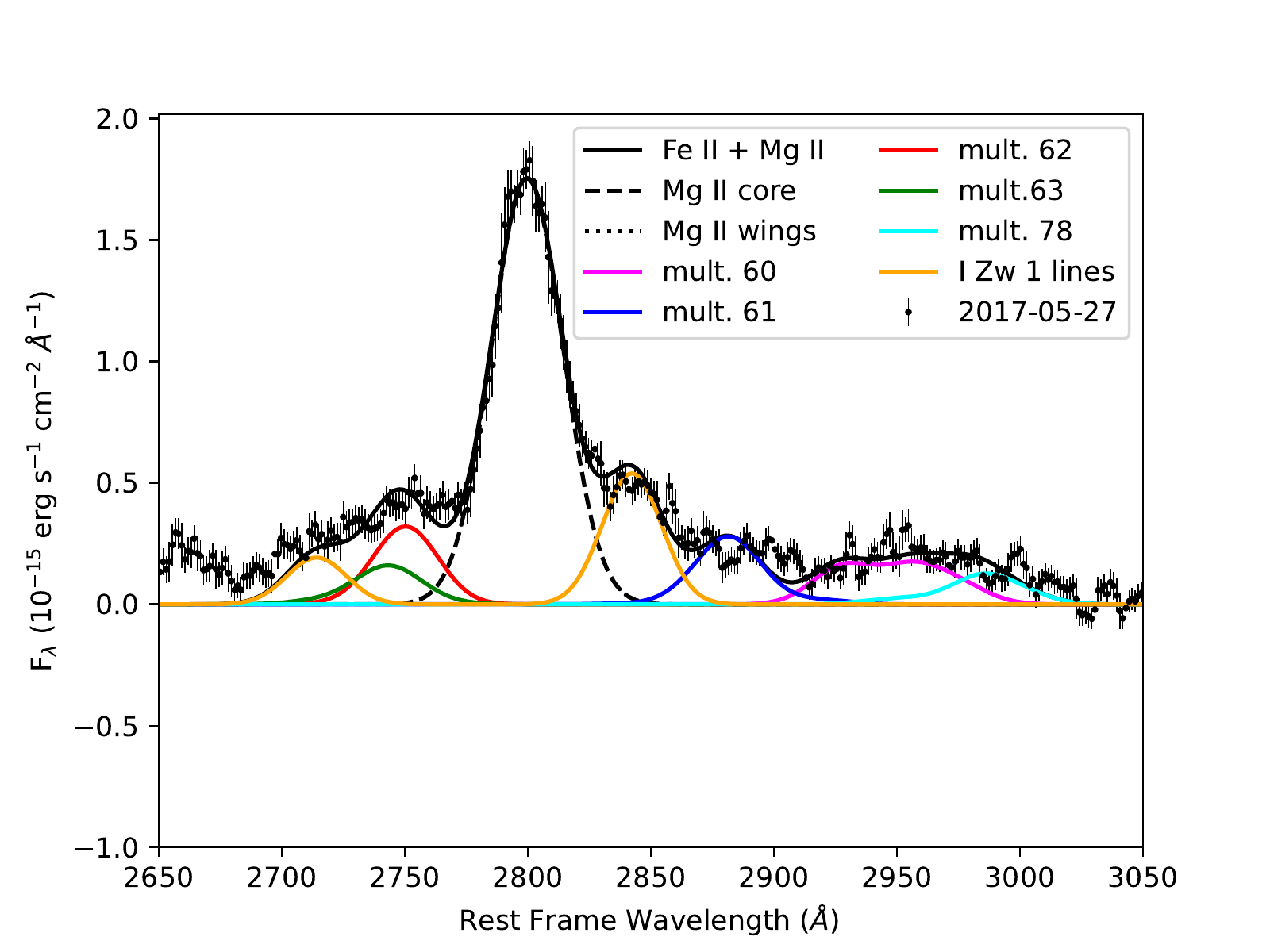}{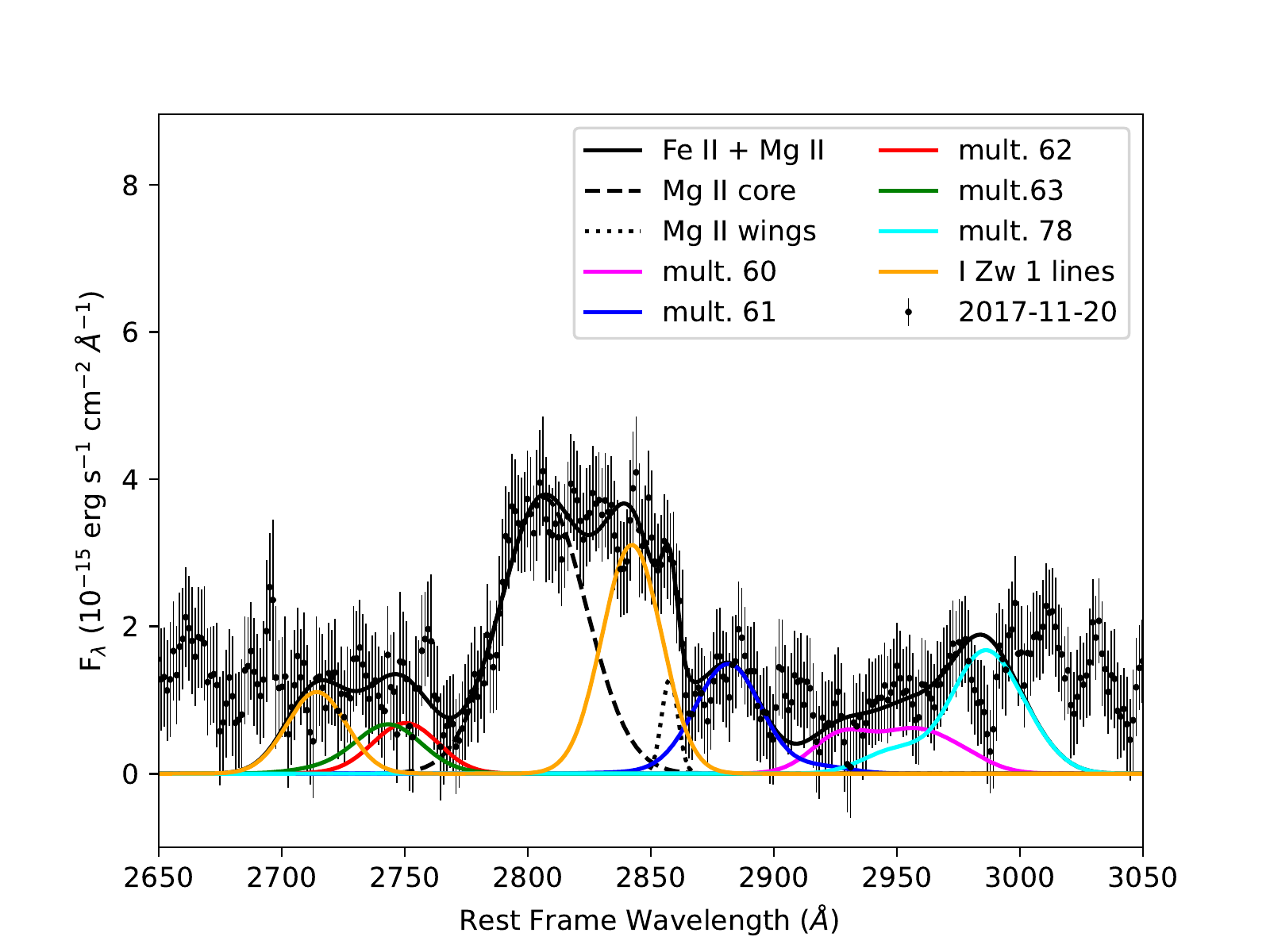}}\\
    {\plottwo{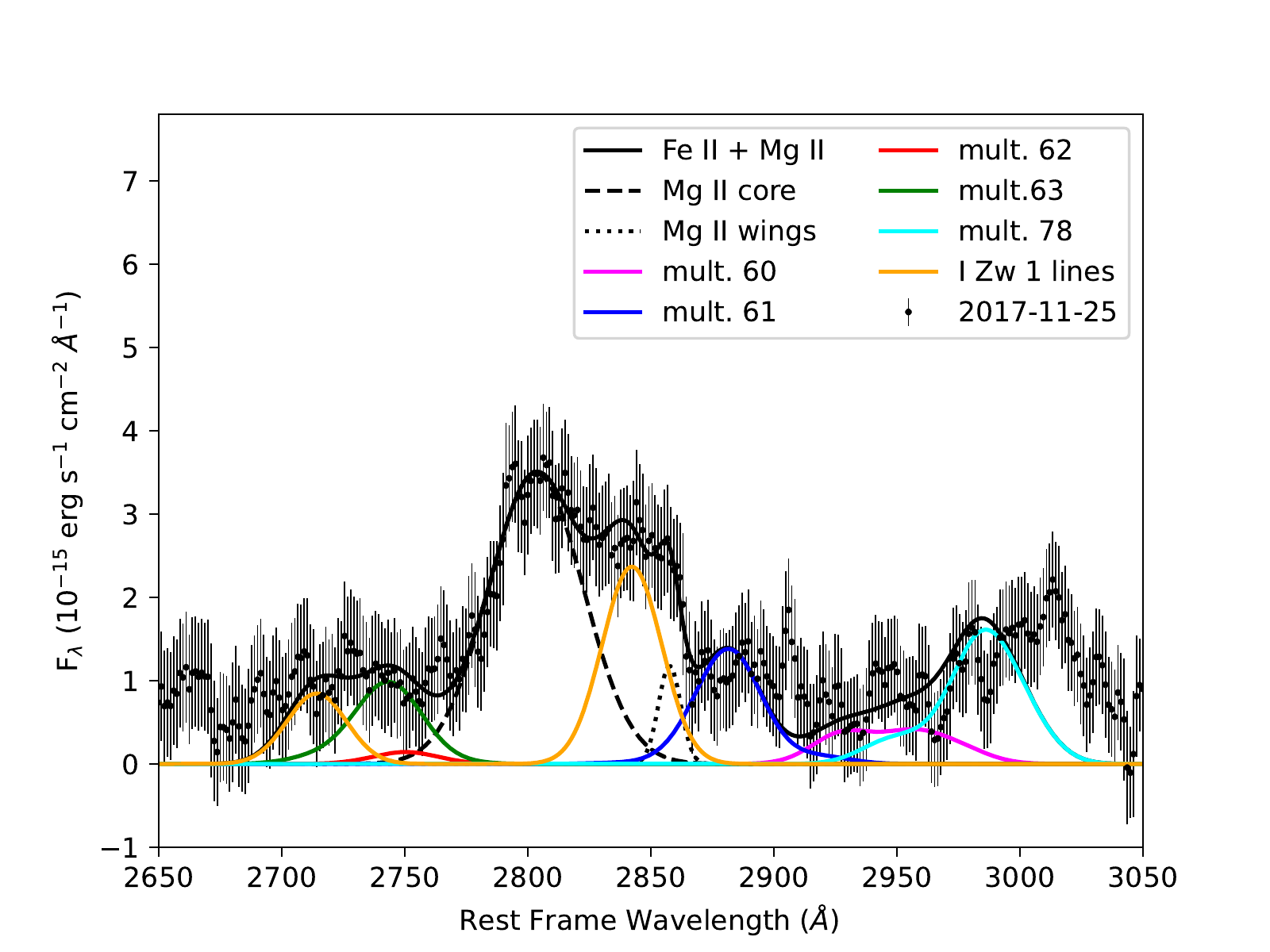}{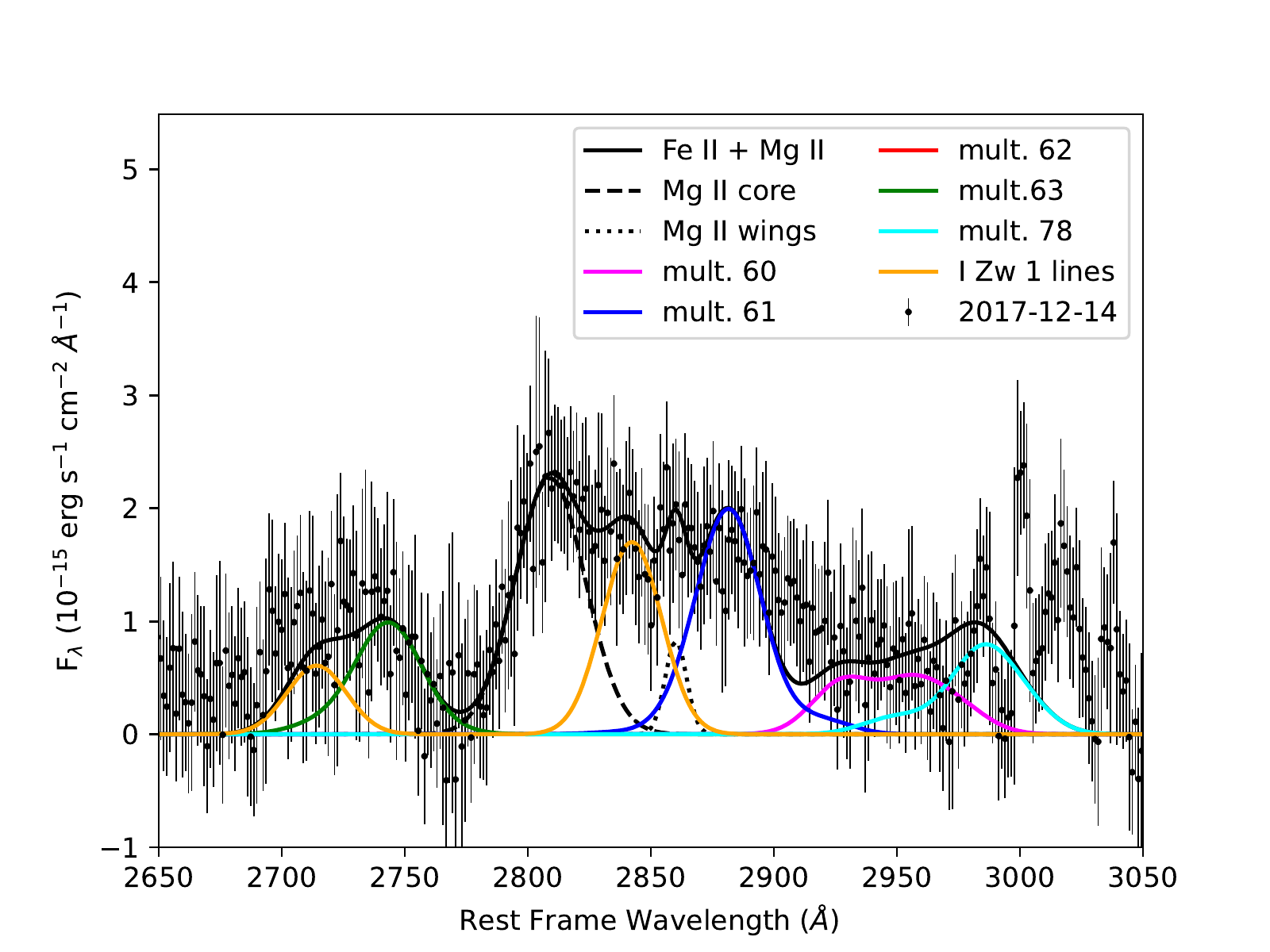}}\\
    {\plottwo{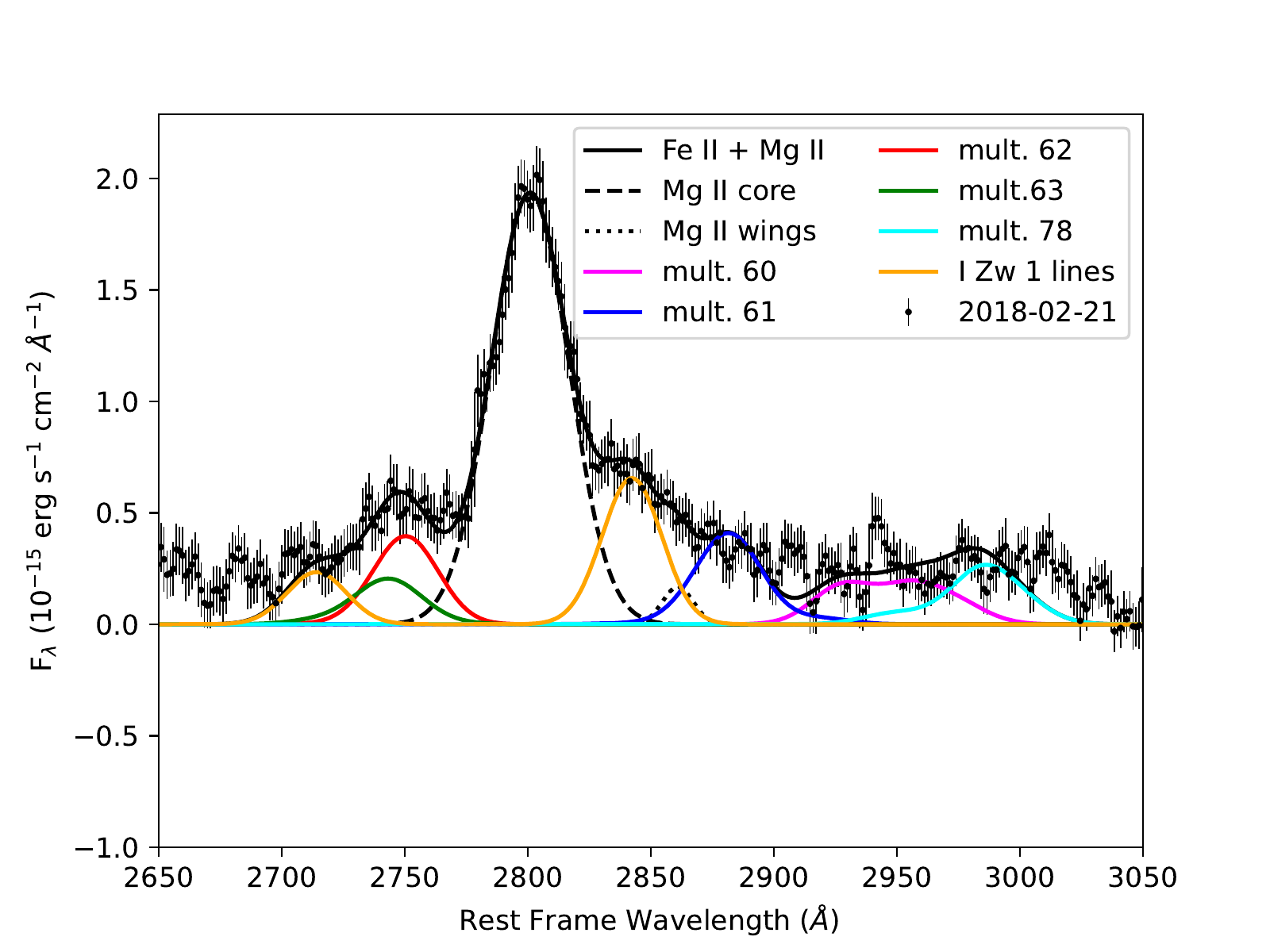}{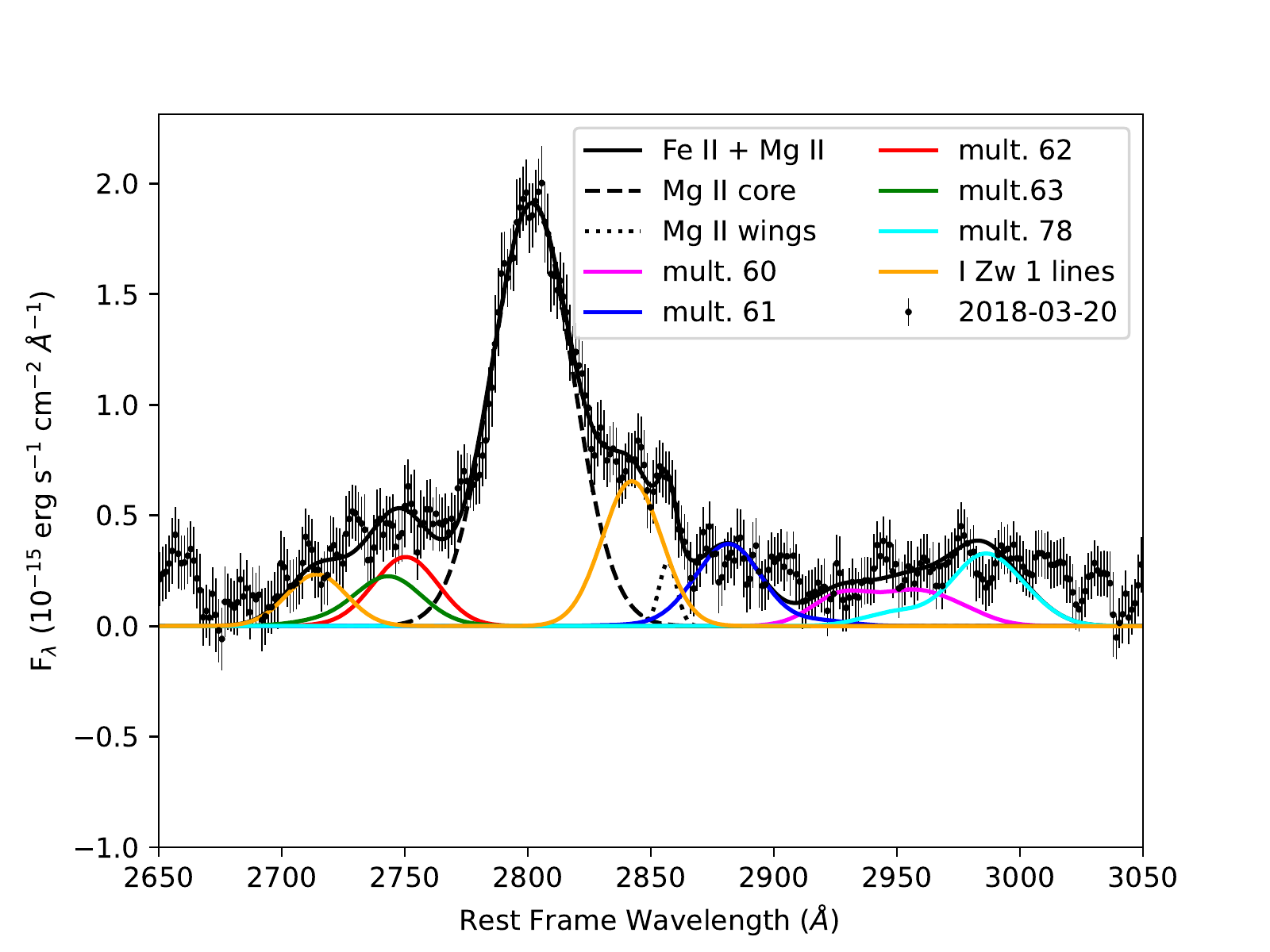}}
    \caption{\ion{Fe}{2} and continuum-subtracted spectra plotted with the best-fit Gaussian \ion{Mg}{2} $\lambda2798$ emission line profile and the scaled Fe II template from \cite{popovic2018}.}
\end{figure*}
\begin{figure*}[!tp]
    \ContinuedFloat
    \centering
    {\plottwo{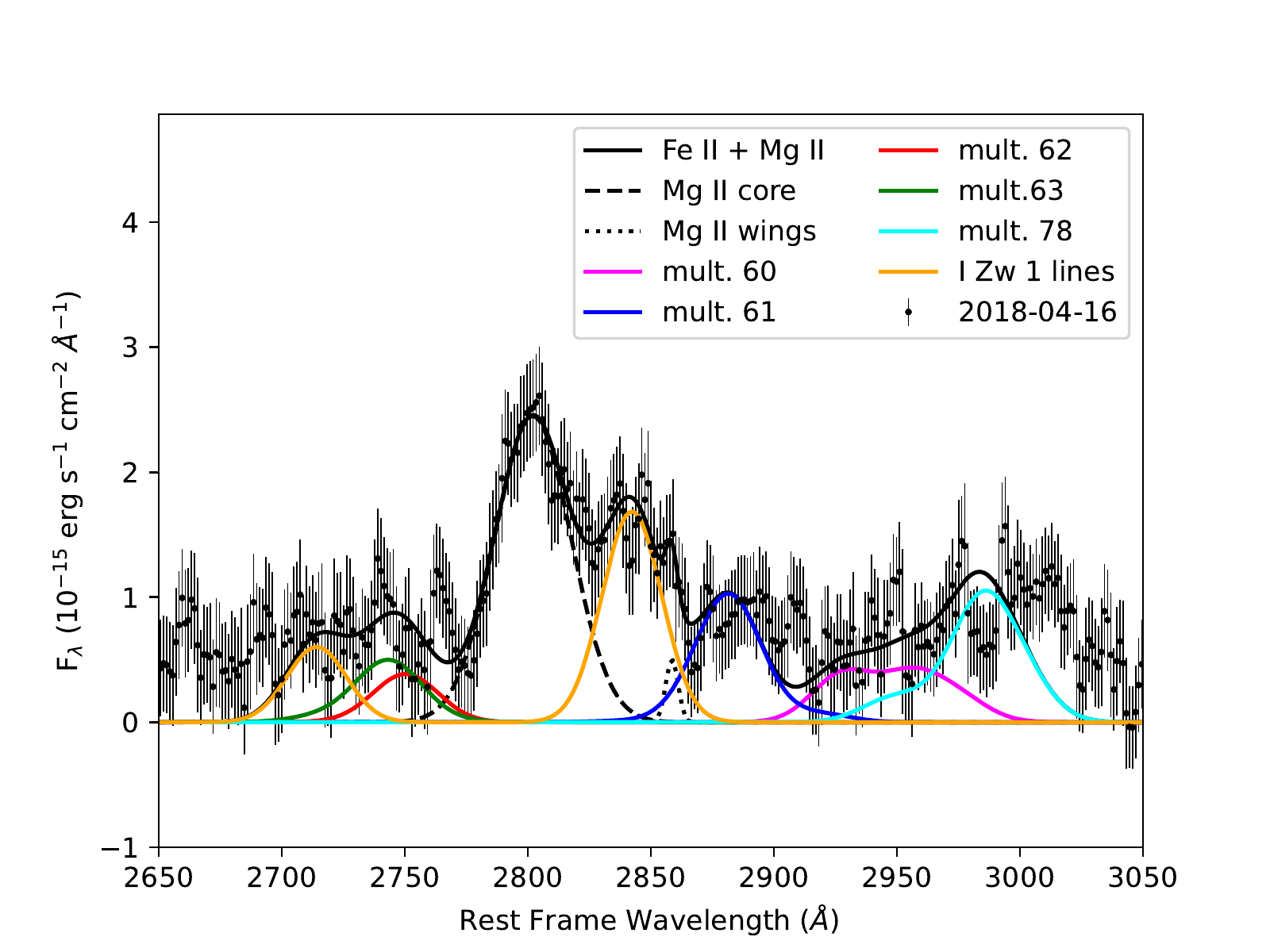}{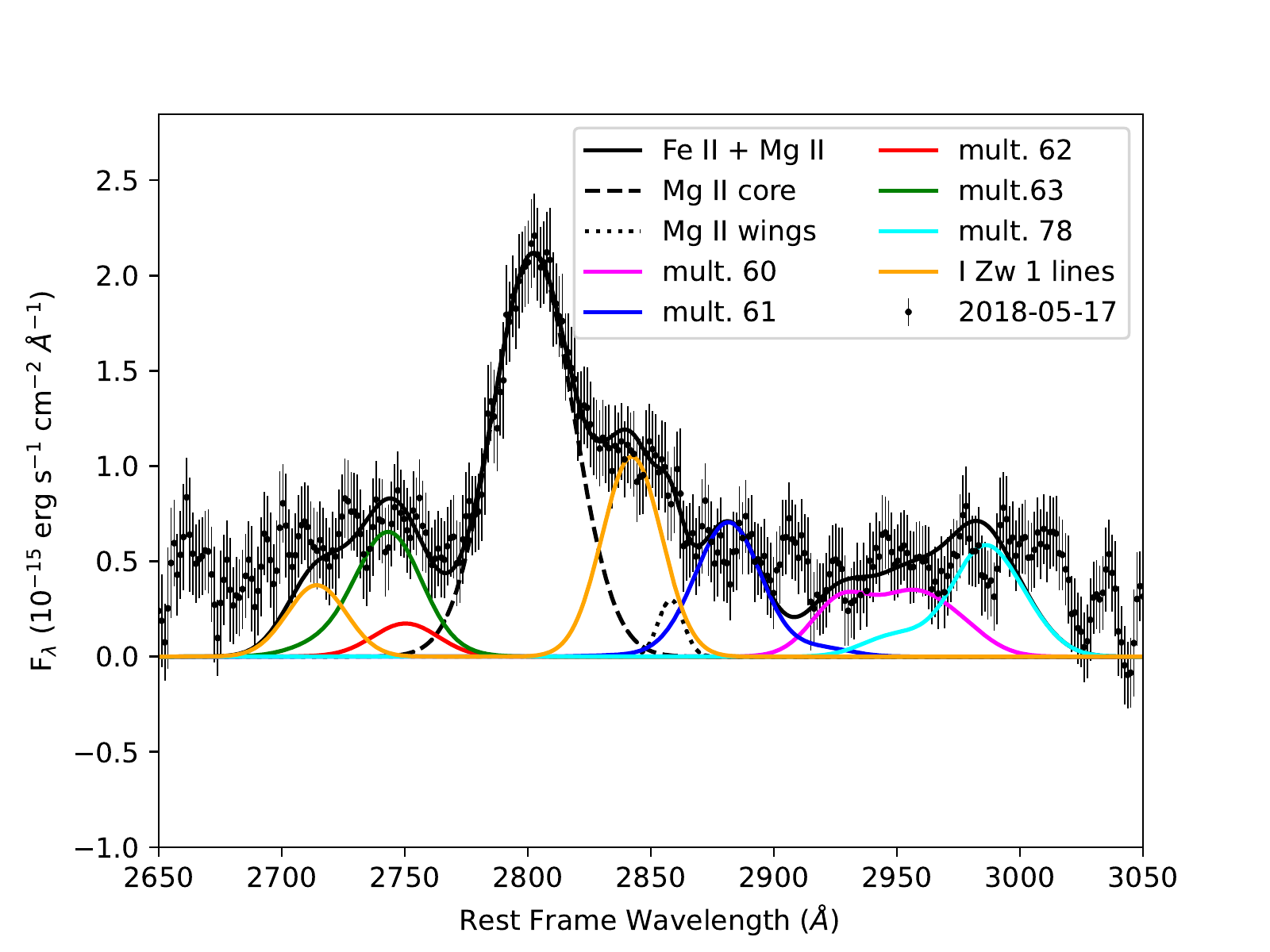}}\\
    {\plottwo{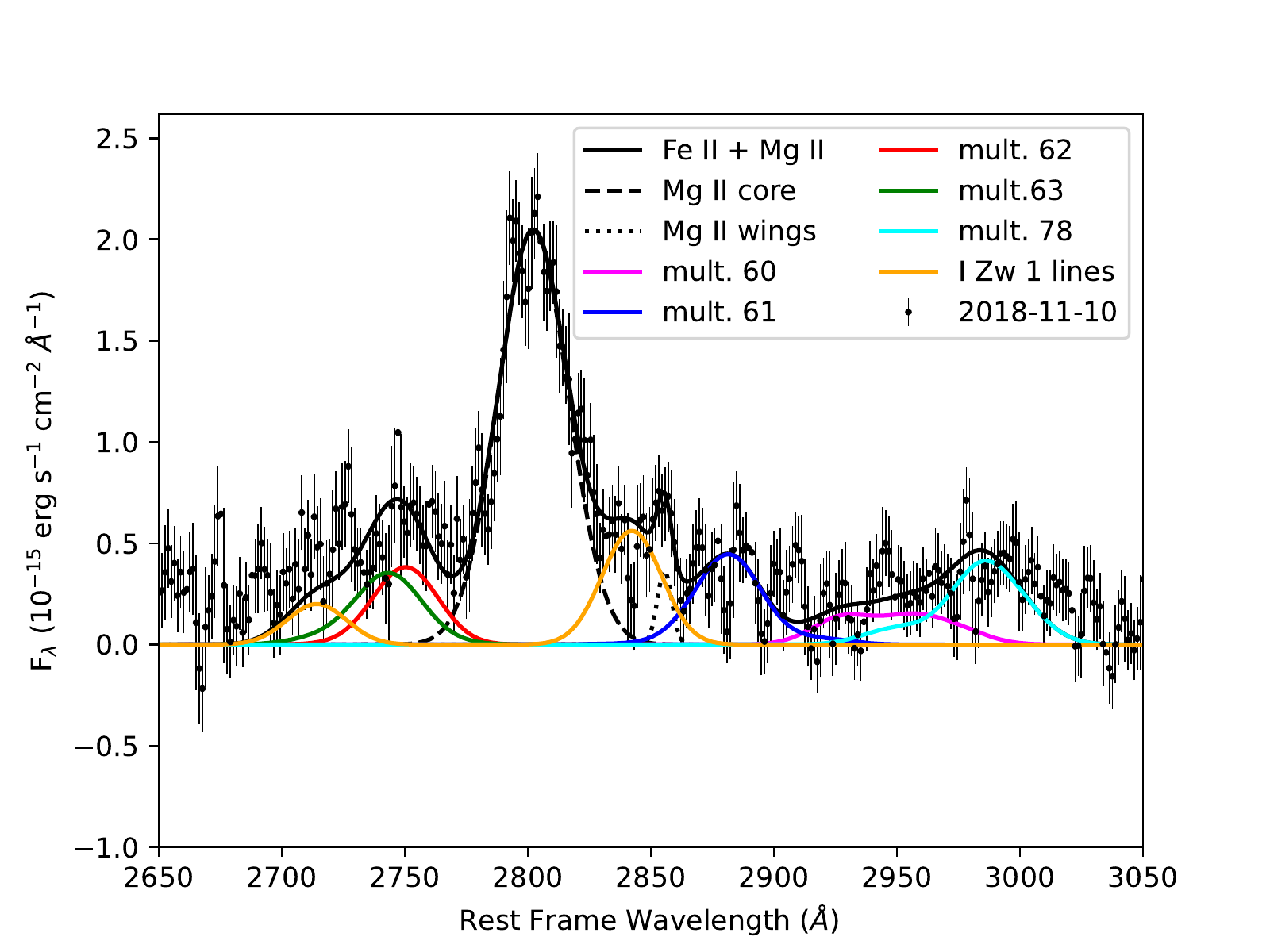}{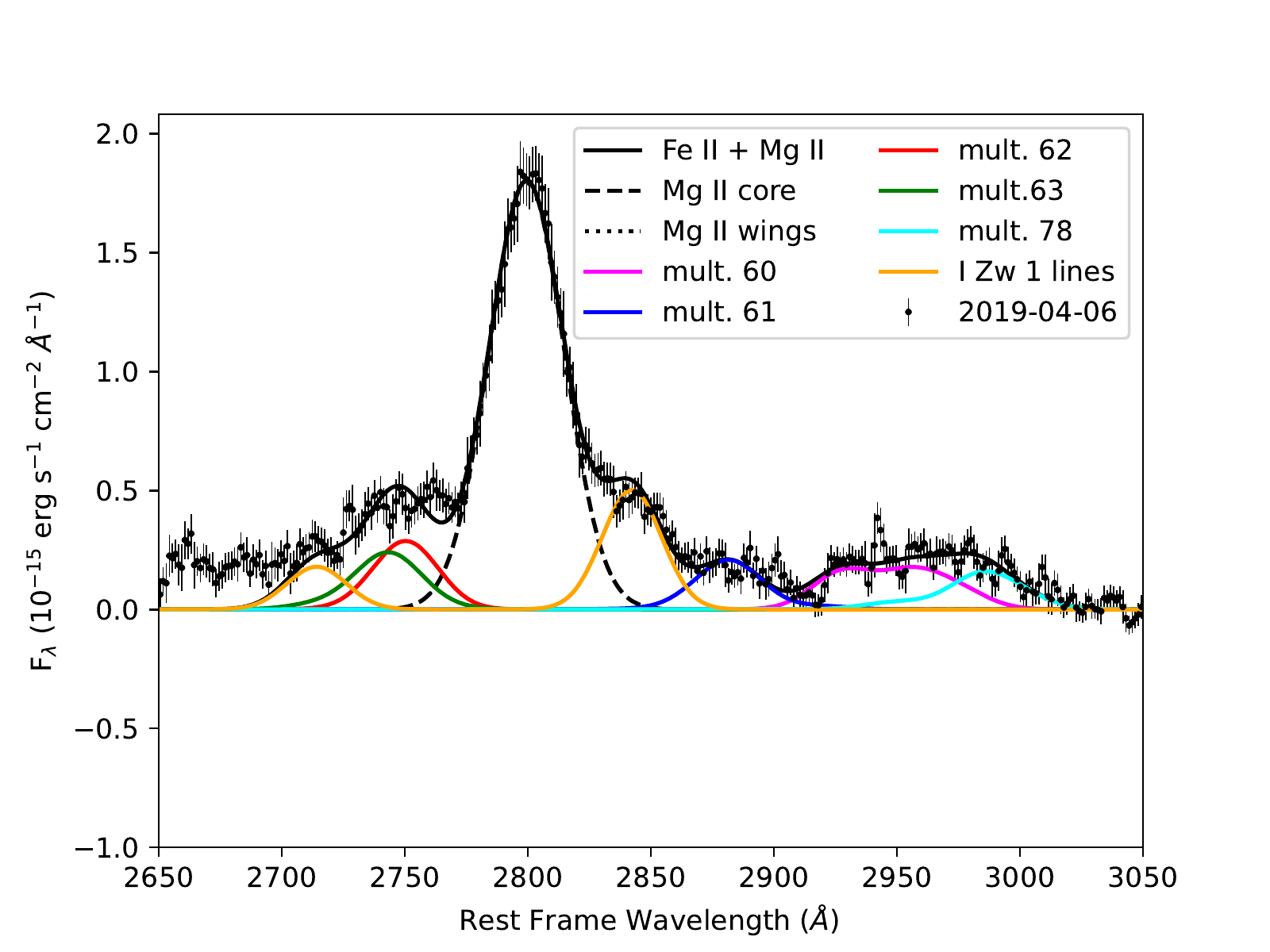}}\\
    \epsscale{0.5}
    {\plotone{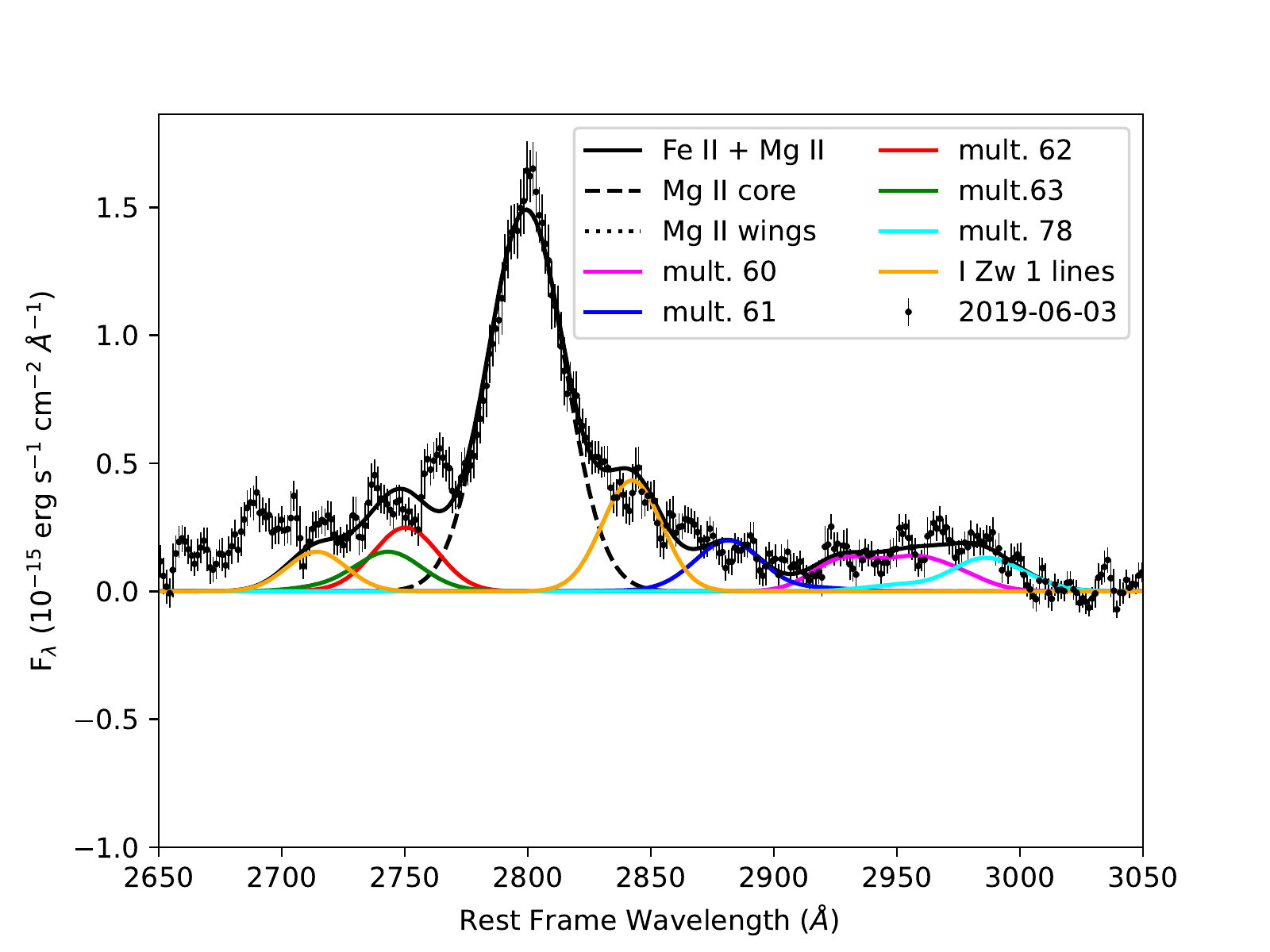}}\\
    \caption{Continued.}
    \label{fig11}
\end{figure*}

\begin{figure*}[!htb]
    \plottwo{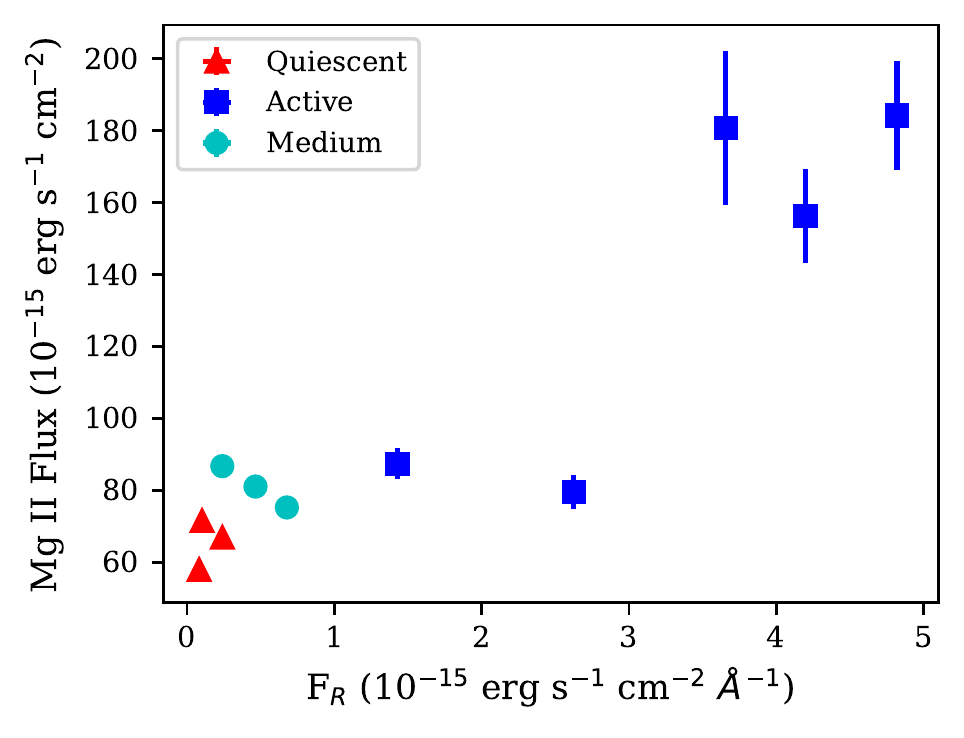}{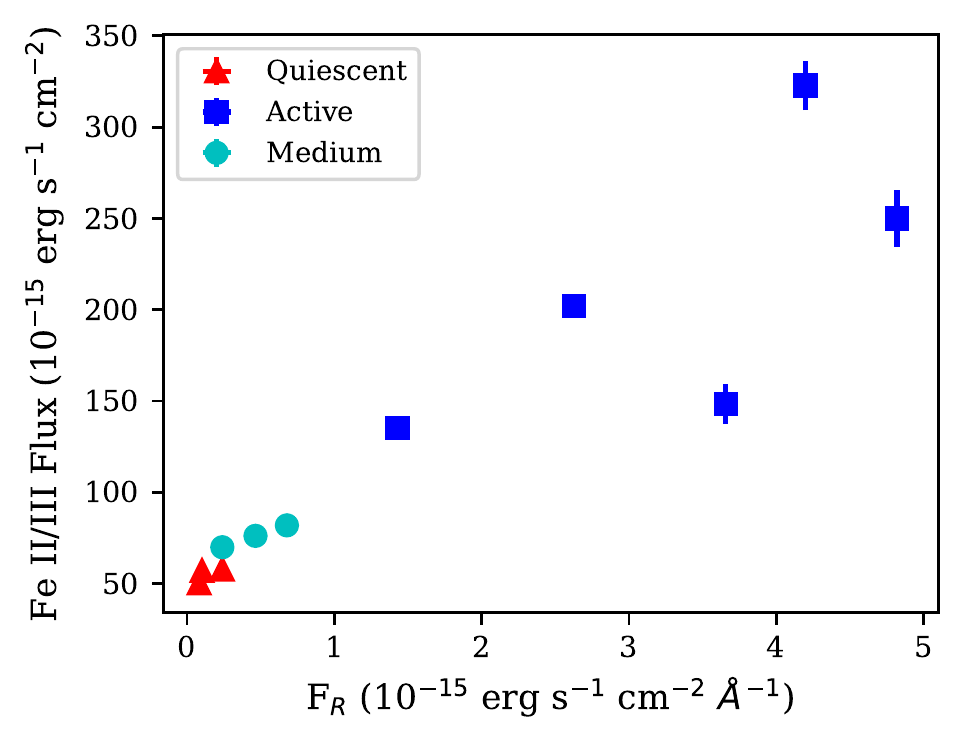}
    \caption{Emission line fluxes of \ion{Mg}{2} (\emph{left}) and \ion{Fe}{2} (\emph{right}) when the Fe template from \cite{popovic2018} was used versus the R band flux density.}
    \label{fig12}
\end{figure*}

\clearpage
\bibliographystyle{aasjournal}
\bibliography{references}

\end{document}